\documentclass{article}
\usepackage{amsmath}
\usepackage{amsfonts}
\usepackage{amssymb}
\usepackage[dvips]{color}
\usepackage{graphicx}%
\newtheorem{Theorem}{Theorem}[section]

\newtheorem{thm}[Theorem]{Theorem}
\newtheorem{dfn}[Theorem]{Definition}

\newenvironment{proof}[1][Proof]{\noindent\textbf{#1.} }{\ \rule{0.5em}{0.5em}}
\newcommand{\bpartial}{\mathop{\partial\kern -4pt\raisebox{.8pt}{$|$}}}
\newcommand{\ket}{\mathclose{]\kern-1.5pt]}}
\newcommand{\bbra}{\mathopen{[\kern-2.2pt[\kern-2.3pt[}}
\newcommand{\bket}{\mathclose{]\kern-2.1pt]\kern-2.3pt]}}

\makeindex
\makeatletter
\newcommand*{\rom}[1]{\expandafter\@slowromancap\romannumeral #1@}
\makeatother

\begin{document}

\title {\large{\bf
  Jacobi structures on real two- and three-dimensional Lie  groups  and their
Jacobi--Lie systems
 }}
\vspace{3mm}
\author { \small{ \bf H.Amirzadeh-Fard$^1$ }\hspace{-2mm}{ \footnote{ e-mail: h.amirzadehfard@azaruniv.ac.ir}},
 \small{ \bf  Gh. Haghighatdoost$^1$}\hspace{-1mm}{ \footnote{ e-mail: gorbanali@azaruniv.ac.ir}},
  \small{ \bf P.
Kheradmandynia$^2$ }\hspace{-1mm}{}
  \small{ \bf A.
Rezaei-Aghdam$^2$ }\hspace{-1mm}{\footnote{ e-mail: rezaei-a@azaruniv.ac.ir }}
\\
{\small{$^{1}$\em Department
of Mathematics,Azarbaijan Shahid Madani University, 53714-161, Tabriz, Iran}}\\
{\small{$^{2}$\em Department of Physics, Azarbaijan Shahid Madani
University, 53714-161, Tabriz, Iran}}\\
  }
 \maketitle
\begin{abstract}
Using the adjoint representations
of Lie algebras,
we classify all  Jacobi structures on real
 two- and three-dimensional
 Lie  groups. Also,
  we study Jacobi--Lie systems on
  these
  real low-dimensional Lie  groups.
Our results are illustrated through examples of Jacobi--Lie Hamiltonian systems on some  real two- and three-dimensional Lie  groups.
\end{abstract}

\smallskip

{\bf keywords:}{\;Lie  group, Jacobi structure, Lie system,  Jacobi--Lie  system.}

\section {\large {\bf Introduction}}

A Lie system is a system  of  $t$-dependent first-order ordinary
 differential equations that describes the integral curves of a  $t$-dependent vector field taking values in a finite-dimensional Lie algebra
of vector fields,  called   Vessiot-Guldberg Lie algebra
\textcolor{red}{ [\ref{ref8}, \ref{LS5}]}.

 Sophus Lie laid down  the study of systems of
first-order ordinary differential equations with a superposition rule  \textcolor{red}{[\ref{ref8}]}.
 In other words,
 a map expressing its general solution in terms of a
generic finite family of particular solutions and some constants.
Systems of first-order differential equations admitting a superposition rule are called Lie systems.



In recent years, many authors have  devoted a great deal of attention to  Lie systems admitting a Vessiot-Guldberg Lie
algebra of Hamiltonian vector fields with respect to a geometric structure
\textcolor{red}{ [\ref{LS5}, \ref{LS1}, \ref{LS2}, \ref{LS3}, \ref{LS4},
\ref{LS6}, \ref{LS7}, \ref{LS8}, \ref{LS9}, \ref{LS10}, \ref{ref5}]}.
In \textcolor{red}{[\ref{ref9}]}, a special case of Lie systems
on  Jacobi manifolds, called Jacobi--Lie systems, admitting
Vessiot-Guldberg Lie algebra
of Hamiltonian vector fields relative to a   Jacobi  structure
 presented
 and  the authors classified Jacobi--Lie systems on the real line and the plane.
In our previous work \textcolor{red}{[\ref{Amirzadeh}]},
we have  studied Jacobi--Lie Hamiltonian systems
on
 real low-dimensional
Jacobi--Lie  groups  and  their
Lie symmetries.

  In this work, we give a method to classify  Jacobi
structures on a Lie group by means of the adjoint representation of the Lie algebra and
classify all  Jacobi structures on real
 two- and three-dimensional
 Lie  groups.
Also, using these Jacobi structures,
we obtain some
 examples of
Jacobi--Lie Hamiltonian systems  on  these  Lie  groups .

The plan of the paper is as follows.  In the second  section, we
 briefly review some definitions of Lie  system and Jacobi--Lie  Hamiltonian system.
 In  section  three,
we describe a useful method for creating Jacobi structures on  low-dimensional Lie groups by using the adjoint representations,
and we calculate these structures on real
 two- and three-dimensional
Lie groups.
 In the fourth  section, using \textcolor{red}{ [\ref{ref9}, \ref{Lee} ]} we obtain
  Jacobi--Lie systems on real two- and three-dimensional  Lie groups.

  \section{   A brief review of Lie  and Jacobi--Lie  Hamiltonian system  }
  For the purpose of self-containment of the paper, we  take a brief review of the  Lie-system  \textcolor{red}{[\ref{ref5}]} and Jacobi--Lie  Hamiltonian system (see the review in   \textcolor{red}{[ \ref{ref9}])}.

\subsection{A  time-dependent vector field and Lie  system  }
  Given two subsets $\mathfrak{a} ,\mathfrak{b}$
  of Lie algebra $\mathfrak{g},$
   we represent by  $[\mathfrak{a} ,\mathfrak{b}]$ the  vector
space spanned by the Lie brackets between elements of $\mathfrak{a}$ and $\mathfrak{b},$
 respectively,
 and  we denote by $Lie(\mathfrak{a})$  the smallest Lie subalgebra of
$(\mathfrak{g} , [., .])$  containing $\mathfrak{a}.$

A time-dependent vector field on a manifold  $M$ is a continuous map
$
X:  \mathbb{R} \times M\longrightarrow TM
$
such that
$ X(t, x)\in T_pM$
for each $(t, x) \in \mathbb{R} \times M.$
In other words,
every time-dependent vector field amounts to a family of vector fields $\lbrace X_t \rbrace _{t\in \mathbb{R}}, $ where
 the map
$
X _{ t}:   M\longrightarrow TM, \,  x\mapsto X_{t} (x)=X(t, x)$
is a vector field on M
and vice versa [\textcolor{red} {\ref{Lee}]}.

An integral curve of a  time-dependent vector field  is an
 integral curve
$\gamma:  \mathbb{R}\longrightarrow \mathbb{R} \times M, \,  t\mapsto (t, x(t)),$
of the suspension of time-dependent vector field \textcolor{red} { [\ref{ref6}]}, namely
\begin{equation}\label{2.1}
\bar{X}:  \mathbb{R} \times M\longrightarrow T(\mathbb{R} \times M)\simeq T\mathbb{R}\oplus TM, \qquad (t, x)\mapsto\dfrac{\partial}{\partial t}+X(t, x).
\end{equation}
 For every
 integral curve
 $\gamma$,
 we have
$\dfrac{dx(t)}{dt}=(X\circ\gamma)(t).$

The smallest Lie  subalgebra of a time-dependent vector field $ X$ on $M$ is the smallest
real Lie algebra containing the  vector fields $\lbrace X_{t}\rbrace _{t\in \mathbb{R}},$ that is,
$\mathfrak{g}^X=Lie (\lbrace X_{t}\rbrace _{t\in \mathbb{R}} )$.

\begin{dfn}
A Lie system is a  time-dependent vector field
$X$
on $M$
whose
$Lie (\lbrace X_{t}\rbrace _{t\in \mathbb{R}} )$
is finite-dimensional \textcolor{red}{[\ref{ref5}]}.
\end{dfn}

\begin{dfn}\label{2.4}
A superposition rule
depending on n particular solutions
 for a time-dependent vector field $ X$ on  $M$ is a function $\Gamma : M^{n} \times M \longrightarrow M,$
 $(x_{(1)}, . . . , x_{(n)};\theta)\longmapsto  x,$
 such  that the general solution $x(t)$ of $X$
 can be brought into the form
$
x(t) = \Gamma(x_{(1)}(t), . . . , x_{(n)}(t); \theta),
$
where $x_{(1)}(t), . . . , x_{(n)}(t) $ is any generic
family
 of particular solutions and $ \theta =
(\theta_1, . . . , \theta_m)$
is a point of M to
be related to initial conditions
 \textcolor{red}{[\ref{ref8}]}.
\end{dfn}
\begin{thm}\label{th2.2}
 A  time-dependent vector field $ X$ on  $M$ admits a superposition
rule if and only if
it has  the form
$
X(t, x)=\sum^r_{i=1} a_i(t)X_i( x)
$
for a certain family $a_1(t), . . . , a_r(t)$ of  time-dependent functions and a family of vector fields $X_1, . . . , X_r $ on
M spanning an r-dimensional real Lie algebra
\textcolor{red}{ [\ref{ref8},\ref{ref11111}]}.
\end{thm}

\subsection{ Jacobi--Lie  Hamiltonian system  }

The notion of
Jacobi manifolds was introduced by
A. Lichnerowicz \textcolor{red}{[\ref{ref4}]}
  and
  A. Kirillov  \textcolor{red}{[\ref{ref3}]}.
   Jacobi manifolds that we want to discuss  are Lichnerowicz's Jacobi
manifolds, known also
 as a local Lie algebra structure on $C^\infty(M, \mathbb{R})$ appeared in the works of  A. Kirillov \textcolor{red}{[\ref{ref3}]}.
Also,
Jacobi--Lie Hamiltonian systems were introduced by Herranz,  Lucas and
Sard\'on
 in \textcolor{red}{[\ref{ref9}]}.

 \begin{dfn}\label{df1}
A Jacobi manifold is a triple  $(M,\mathbf{ \Lambda, E)}$ where $\mathbf{\Lambda} \in\Gamma(\large{\bigwedge}^2 TM)$ and
 $\mathbf{E} \in\Gamma(TM)$
 satisfying
\begin{equation}\label{Jacobi}
[[\mathbf{\Lambda, \Lambda}]]=2\mathbf{E}\wedge \mathbf{ \Lambda},\qquad [[\mathbf{E, \Lambda}]]=0,
\end{equation}
\end{dfn}

where
$[[., .]]$ stands for the
 Schouten-Nijenhius bracket (see  \textcolor{red}{[\ref{Vaisman}]} for details).
\begin{dfn}

A  vector field
$X_f$
 on a Jacobi manifold $(M,\mathbf{ \Lambda, E)}$
is said to be  a Hamiltonian vector field
if
it can
be written in the form
\begin{equation}\label{vector}
X_f= [[\mathbf{ \Lambda}, f]]+f\mathbf{ E}={\mathbf{\Lambda}}^{\#}(d f)+f\mathbf{ E}
\end{equation}
for   $f \in C^{\infty}( M)$  which is called the Hamiltonian
\textcolor{red}{[\ref{ref9}]}.
\end{dfn}

\begin{dfn}\label{}
A Lie system is said to be a
Jacobi--Lie system
  if it admits
a Vessiot-Guldberg Lie algebra
$\mathfrak{g}$
of Hamiltonian vector fields with respect
to a Jacobi structure \textcolor{red}{[\ref{ref9}]}.
 \end{dfn}

 \begin{dfn}\label{}
 A Jacobi--Lie system is said to be
a Jacobi--Lie  Hamiltonian system $(M, \Lambda, E, f)$
if $ X_{f_t}$  is a Hamiltonian vector field with Hamiltonian function $f_t$
such that $Lie (\lbrace  f_{t} \rbrace_{t\in  \mathbb{R} }, \lbrace ., .\rbrace_{\Lambda,E}) $ is finite-dimensional.
Where
$ f:  \mathbb{R}     \times M \longrightarrow M, \,
 (t, x) \mapsto f_{t} (x)$
 is a
t-dependent function
$\forall t \in \mathbb{R}$
\textcolor{red}{[\ref{ref9}]}.
\end{dfn}

\section{Classification of Jacobi structures on real low-dimensional Lie groups   }

 Here, using non-coordinate bases, we translate the Jacobi structures on a Lie group into a  Lie algebra.
 Next, applying the adjoint representation
for the Lie algebras and the vielbeins for Lie group, we classify Jacobi structures on real two and three dimensional  Lie groups.
\subsection{  Jacobi   structures on manifolds }



 Let $x^{\mu}(\mu=1, . . . ,dimM)$
 be the local coordinates of a manifold  $M$.
We have the following relations
for  the Jacobi structure
$(\mathbf{ \Lambda, E)}$ on $M$:
\begin{equation}\label{}
\mathbf{\Lambda } = \frac{1}{2}{\mathbf{\Lambda} ^{\mu\nu}}\partial_ {\mu}\wedge \partial_ {\nu},
\end{equation}
\begin{equation}\label{}
\mathbf{ E }= \mathbf{E}^{\mu} \partial _{\mu},
\end{equation}
and the Jacobi bracket  on $M$:
\begin{equation}\label{s4}
\{ f,g\}_{\mathbf{ \Lambda, E} } = {\mathbf{\Lambda} ^{\mu\nu}}\partial_ {\mu}f\partial_ {\nu}g + f{\mathbf{ E}}^{\mu}\partial_ {\mu} g - g\mathbf{ E}^{\mu}\partial _{\mu}f,\quad \forall f,g \in C^{\infty}(M),
\end{equation}
where
 we use  Einstein's summation
convention, and
 the symbol
$\partial _{\mu}f$
means that we compute the usual derivative of  $f$ with respect to $x^{\mu}$.

Substituting   the Jacobi bracket (\ref{s4}) in the  Jacobi identity, one can  obtain the following relations
\begin{equation}\label{s5}
{\mathbf{\Lambda} ^{\nu \rho }}{\partial _\rho }{\mathbf{\Lambda} ^{\lambda \mu } + \mathbf{\Lambda} ^{\mu \rho }{\partial _\rho }{\mathbf{\Lambda} ^{\nu \mathbf{\lambda} }} + {\mathbf{\Lambda }^{\mathbf{\lambda} \rho }}{\partial _\rho }{\mathbf{\Lambda }^{\mu \nu }} + \mathbf{E}^\mathbf{\lambda} }{\mathbf{\Lambda }^{\mu \nu }} + \mathbf{E}^{\mu }{\mathbf{\Lambda} ^{\nu \mathbf{\lambda }}} + \mathbf{ E}^{\nu }{\mathbf{\Lambda} ^{\mathbf{\lambda} \mu }} = 0,
\end{equation}
\begin{equation}\label{s6}
\mathbf{E}^{\rho }{\partial _\rho }{\mathbf{\Lambda} ^{\mu \nu }} - {\mathbf{\Lambda} ^{\rho \vartheta }}{\partial _\rho }\mathbf{ E}^{\mu } + {\mathbf{\Lambda }^{\rho \mu }}{\partial _\rho }\mathbf{ E}^{\nu } = 0.
\end{equation}

 The equations (\ref{s5}) and (\ref{s6} ) have called the Jacobi equations
 which can also be obtain from
$(\ref{Jacobi}). $
To obtain the general form of the Jacobi structures on a manifold $M,$
one can compute
the general solution
for the Jacobi equations
\textcolor{red}{ [\ref{hass}]}.


As a Lie group is a smooth manifold  and it is also a group in the
algebraic sense, we have the Jacobi structures on  Lie group.

\subsection{  Jacobi   structures on   real  low-dimensional Lie algebras  }

 Now using non-coordinate bases, we  translate   the equations (\ref{s5}), (\ref{s6}) from a Lie group  into a  Lie algebra.

\begin{dfn}\textcolor{red}{ [\ref{naka}]}
In the coordinate basis, $T_p M$ is spanned by $\{ {e_\mu }\}  = \{ {\partial _\mu }\} $ and $T_p ^*M $ by
$\{ d{x^\mu }\} $.
Considering their linear combination,
we have
\begin{equation}
\hat{e}_a= e_{a}^{\;\;\mu} \; \partial _\mu,\qquad \hat{\theta}^{a}= e_{\;\;\mu}^{a } dx^{\mu},\qquad
\lbrace e_{a}^{\;\;\mu}\rbrace\in GL(m,R),
\end{equation}
where
det  $e_{a}^{\;\;\mu}>0$
; that is to say, $\hat{e}_a$ is the frame of basis vectors which is
obtained by a $ GL(m,R)$-rotation of the basis $\{ {e_\mu }\}$ preserving the orientation.

Applying
$\langle\hat{e}_a, \hat{\theta}^{b}\rangle=\delta^{\;\;b}_a$,
we obtain
\begin{equation}
e_{\;\;\mu}^{a } e_{a}^{\;\;\nu} =\delta_{\mu} ^{\;\;\nu},\qquad  e_{\;\;\mu}^{a } e_{b}^{\;\;\mu} =\delta_{\;\;b} ^{a},
\end{equation}
where  $e_{\;\;\mu}^{a }$is inverse of $e_{a}^{\;\;\mu},$ and  indices $ \mu, \nu, \cdots $ and  $a, b, \cdots$ related to the coordinates of the  Lie group  and the basis of the Lie algebra, respectively.
The bases $\lbrace\hat{e}_a\rbrace$ and $\lbrace\hat{\theta}^{a}\rbrace$ are called the non-coordinate bases \textcolor{red}{ [\ref{naka}]}.

Using   the  $\hat{e}_a= e_{a}^{\;\;\mu} \; \partial _\mu$, we have
\begin{equation}
[\hat{e}_a, \hat{e}_b]=\mathbf{f}_{ab}^{\;\;c}\hat{e}_c,
\end{equation}
where the coefficient
$\mathbf{f}_{ab}^{\;\;c}$
related to the vielbein
$ e_{a}^{\;\;\mu}$
with the following
Maurer-Cartan
relation:
\begin{equation}\label{Maurer-Cartan}
{\mathbf{f}}_{ab}^{\;\;c}=e^c_{\;\;\nu}( e_{a}^{\;\;\mu}\partial _\mu e_{b}^{\;\;\nu}-e_{b}^{\;\;\mu}\partial _\mu e_{a}^{\;\;\nu}).
\end{equation}
These coefficients are the structure constants of the Lie algebra
 $\mathfrak{g}$
  of the Lie group
  $\mathbf{G}$
  when $M$ is a Lie group
  $\mathbf{G}$
 \textcolor{red}{[\ref{naka}]}.
\end{dfn}

 Writing the Jacobi structure
 $(\mathbf{G, \Lambda, E)}$
 in terms of the non-coordinate basis, we have
\begin{equation}\label{basis}
{\mathbf{\Lambda }^{\mu \nu }}= e_{a}^{\;\;\mu} e_{b}^{\;\;\nu} {\Lambda ^{ab }},
\end{equation}
\begin{equation}\label{basis2}
\mathbf{ E}^{\mu}= e_{a}^{\;\;\mu}E^{a},
\end{equation}
where
the  Jacobi   structure
 $\Lambda^{ab} $  and $E^a$  are related to Lie algebra
 and
 we have assumed that these are independent
of the coordinate of Lie group.

Inserting
(\ref{basis})  and (\ref{basis2}) into (\ref{s5}) and (\ref{s6}), and using the Maurer-Cartan equation
(\ref{Maurer-Cartan})
one can obtain
 the following relations:
\begin{equation}\label{lie 17}
{{\mathbf{f}}_{bc}}^f{\Lambda ^{hb}}{\Lambda ^{ce}} + {{\mathbf{f}}_{bd}}^e{\Lambda ^{hb}}{\Lambda ^{fd}} + {{\mathbf{f}}_{ba}}^h{\Lambda ^{eb}}{\Lambda ^{af}} + {{\rm E}^f}{\Lambda ^{eh}} + {{\rm E}^e}{\Lambda ^{hf}} + {{\rm E}^h}{\Lambda ^{fe}} = 0,
\end{equation}
\begin{equation}\label{lie 16}
{{\mathbf{f}}_{ac}}^d{{\rm E}^a}{\Lambda ^{ce}} + {{\mathbf{f}}_{ab}}^e{{\rm E}^a}{\Lambda ^{db}} = 0.
\end{equation}


  Working with  tensorial form of the equations  (\ref{lie 17}) and (\ref{lie 16}) is far  difficult; as a result,
 we suggest writing these equations in matrix forms using the following adjoint representations for
Lie algebras

\begin{equation}\label{rep}
{{\mathbf{f}}_{ab}}^c =  - {({\chi _a})_b}^c,\qquad
{{\mathbf{f}}_{ab}}^c =  - {({\cal Y}^c)_{ab}},\qquad
\end{equation}

Hence, the relations   (\ref{lie 17}) and (\ref{lie 16}) can be rewritten as follows, respectively
\begin{equation}\label{Lie al2}
 -\Big ({\Lambda ^{ce}}({\chi ^t}_c\Lambda ) + \Lambda {\cal Y}^e\Lambda  + (\Lambda \chi _b){\Lambda ^{be}} + {{\rm E}^e}\Lambda \Big) ^{fh}+ {{\rm E}^f}{\Lambda ^{eh}} + {\Lambda ^{fe}}{{\rm E}^h} = 0,
\end{equation}
\begin{equation}\label{Lie al1}
(\Lambda \chi _a - {(\Lambda {\chi _a})^t}){{\rm E}^a} = 0.
\end{equation}
To compute
the general solutions
for the equations (\ref{Lie al2}) and (\ref{Lie al1}), we use the  Maple program.
The solutions of equations (\ref{Lie al2}) and (\ref{Lie al1}) yield the Jacobi structures on
real low-dimensional Lie algebras.
 The results
 for
  two- and three-dimensional
 Lie algebras
  have been listed in Tables 3 and 4.


Note that in
the classification of  these  Jacobi structures
 some of these Jacobi structures are equivalent,
so to classify them, we  must define an equivalence relation and use the following theorem.

\begin{thm}\label{}
Two Jacobi structures
$(\Lambda, E)$
 and
$(\Lambda^\prime,E^\prime)$
 are equivalent if there exist
 $A \in Aut(\mathfrak{g}), $
$ ( $ i.e.,
 automorphism group of the Lie algebra
$\mathfrak{g})$
 such that
 \begin{equation}\label{p100}
\Lambda=A^t \Lambda^\prime A,
\end{equation}
 and
  \begin{equation}\label{r100}
E^e=E^{\prime^b} A_b^{\,\,e}.
\end{equation}
\end{thm}\label{}
\begin{proof}
By definition of
 automorphism of the Lie algebra
  $ \mathfrak{g}$
  with the basis
  $\lbrace X_a \rbrace$
  and the structure constants
  $\mathbf{f}_{ab}^{\;\;c} ,$
 $A:  \mathfrak{g}\longrightarrow \mathfrak{g}$
 we have
 \begin{equation}\label{bors1}
AX_a=A_a^{\,\,b}X_b,
\end{equation}
 where
 $A_a^{\,\,b}$
 satisfies the following relation
 \begin{equation}\label{Automor}
A_a^{\,\,k}f_{kl}^{\,\,\,m}A_b^{\,\,l}=f_{ab}^{\,\,\,c}A_c^{\,\,m}.
\end{equation}
Applying
(\ref{rep}) in (\ref{Automor}),
one can obtain  matrix relations
\begin{equation}\label{metaut1}
A{\cal Y}^{m}A^{t}={\cal Y}^{c}A_c^{\,\,m},
\end{equation}
and
\begin{equation}\label{metaut2}
A\chi_{l}A_b^{\,\,l}=\chi_bA,
\end{equation}
where $A$ is
the matrix form of
$A_a^{\,\,b}$
and
$A^{t}$
is the transpose of  $A$.

Inserting ( \ref{p100})
and
( \ref{r100})
  into  ($\ref{Lie al2}$) and using
  $( \ref{metaut1})$
  and
  $( \ref{metaut2}),$
and multiply the
left side of the above equation  by
$(A^{-t})_s^{\,\,f}$
 and the right  side  by
 $(A^{-1})_e^{\,\,l}(A^{-1})_h^{\,\,p}$,
  one can obtain the following relation
\begin{equation*}\label{}
 -  \Big({\Lambda^{\prime ^{ab}}}({\chi ^t}_a\Lambda^{\prime} ) +
  \Lambda^{\prime} {\cal Y}^b\Lambda^{\prime}  +
   (\Lambda^{\prime} \chi _d){\Lambda ^{\prime^{db}}} +
    E^{\prime^b}\Lambda^{\prime} \Big ) ^{su}
 +  E^{\prime^s}{\Lambda ^{\prime^{bu}}} + {\Lambda ^{\prime^{sb}}} E^{\prime^u} = 0.
\end{equation*}
Hence, $(\Lambda, E)$ and $(\Lambda ^{\prime}, E^{\prime})$ are
solutions of the equations
(\ref{Lie al2}) and (\ref{Lie al1}) and they are
equivalent.

In the same way by
inserting
$( \ref{p100})$ and
( \ref{r100}) into  ($\ref{Lie al1}$),
 one can obtain
\begin{equation}\label{}
(A^t \Lambda^\prime A \chi _a - {(A^t \Lambda^\prime A {\chi _a})^t})E^{\prime^b} A_b^{\,a} = 0,
\end{equation}

 then using
$( \ref{metaut2})$,
and multiply the
left side of the equation  by
$A^{-t}$
 and the right side  by
 $A^{-1}$ we have

\begin{equation}\label{}
 (\Lambda^\prime \chi _b - {( \Lambda^\prime  {\chi _b})^t})E^{\prime^b}  = 0.
\end{equation}
So $E$ and $E ^{\prime}$ are equivalent and consequently
   Jacobi structures
$(\Lambda, E)$
 and
$(\Lambda^\prime,E^\prime)$
 are equivalent.
\end{proof}

 For an illustration of the method of  the classification (Tables 3 and 4), here we explain our method for  the three dimensional  Bianchi type Lie  algebra $\rom{3}$.

\subsection{ An example for $\rom{3}$ Lie algebra }

In Tables 1 and 2 we have   used the Landau and Lifshitz
classification \textcolor{red}{[\ref{patera}]} for  three  dimensional Lie algebras,
and we have  applied
 the Patera and Winternitz
classification\textcolor{red}{ [\ref{Landau}]} for  two dimensional Lie algebras (see Tables 1 and 2 ).\\
Let us illustrate
all details of the method of  the classification of  the Jacobi structure on real low-dimensional Lie algebra for the Lie algebra
 $\rom{3}$
with the following non-zero commutator
$$[X_1, X_2]=-(X_2+X_3), \quad [X_1, X_3]=-(X_2+X_3).$$
We write the  matrix form of the bivector field
$\Lambda$
 and  Reeb vector field
 $E$
as follows:
\begin{equation}\label{reeb1}
 \Lambda=\left( {\begin{array}{*{20}{c}}
0&\lambda _{12}  &\lambda _{13}\\
-\lambda _{12} &0 &\lambda _{23}\\
-\lambda _{13} &-\lambda _{23} &0
\end{array}} \right),  \qquad
E=\left( {\begin {array}{c}e_{1}\\ \noalign{\medskip}e_{2}\\ \noalign{\medskip} e_{3} \end {array} }\right),
\end{equation}

where
$\lambda _{ij} $ and   $e_{i}$
can be any arbitrary  real constant.

Using
$(\ref{rep}),$
we obtain
 the following adjoint representations
for Lie algebra  $\rom{3}$.

\begin{equation}\label{reeb2}
\chi_1=\left( \begin {array}{ccc} 0&0&0\\ \noalign{\medskip}0&1&1
\\ \noalign{\medskip}0&1&1\end {array} \right),\qquad
\chi_2=\left( \begin {array}{ccc} 0&-1&-1\\ \noalign{\medskip}0&0&0
\\ \noalign{\medskip}0&0&0\end {array} \right),
\qquad
\chi_3=\left( \begin {array}{ccc} 0&-1&-1\\ \noalign{\medskip}0&0&0
\\ \noalign{\medskip}0&0&0\end {array} \right),
\end{equation}
\begin{equation}\label{reeb3}
{\cal Y}_1= \left(\begin {array}{ccc} 0&0&0\\ \noalign{\medskip}0&0&0
\\ \noalign{\medskip}0&0&0\end {array} \right),
\qquad
{\cal Y}_2=\left( \begin {array}{ccc} 0&1&1\\ \noalign{\medskip}-1&0&0
\\ \noalign{\medskip}-1&0&0\end {array} \right),
\qquad
{\cal Y}_3=\left( \begin {array}{ccc} 0&1&1\\ \noalign{\medskip}-1&0&0
\\ \noalign{\medskip}-1&0&0\end {array} \right),
\end{equation}
Inserting
$( \ref{reeb1})$, ($\ref{reeb2}$) and ($\ref{reeb3}$)  in  ($\ref{Lie al1}$) and ($\ref{Lie al2}$) one can find  the
$\Lambda$ and  $E$ for  the Lie algebra.

One of the solutions has the following forms:

\begin{equation}\label{jac}
\Lambda=
\left( \begin {array}{ccc} 0&\lambda_{{12}}&\lambda_{{13}}
\\ \noalign{\medskip}-\lambda_{{12}}&0&\lambda_{{23}}
\\ \noalign{\medskip}-\lambda_{{13}}&-\lambda_{{23}}&0\end {array}
 \right),\qquad
E=\left( \begin {array}{c} 0\\ \noalign{\medskip} -\lambda_{{13}}+\lambda_{{12}}\\ \noalign{\medskip}\lambda_{{13}}-\lambda_{{12}}
\end {array} \right).
\end{equation}
By the preceding theorem,
we  show that
the Jacobi structure
$(\Lambda, E)$ contains the following equivalence classes:\\


Applying  the following automorphism group  of a Lie algebra
$\rom{3}$
(\textcolor{red}{[\ref{Hemmati}]}, see also Table 5)

 \begin{equation}\label{}
A=
\left( \begin {array}{ccc} 1&a_{12}&a_{13}
\\ \noalign{\medskip}0&a_{22}&a_{23}
\\ \noalign{\medskip}0&a_{23}&a_{22}
\end {array} \right),
\end{equation}
 and inserting the above automorphism in the relations
  \begin{equation*}\label{prime}
\Lambda^\prime=A^t \Lambda A,
\end{equation*}
 and
  \begin{equation*}\label{reb}
E^\prime=E A,
\end{equation*}
we obtain

\begin{equation}
A=\left( \begin {array}{ccc} 1&{\dfrac {\lambda_{{23}}}{{\lambda_{{12}}}
^{2}-{\lambda_{{13}}}^{2}}}&b\\ \noalign{\medskip}0&-{\dfrac {\lambda_{
{13}}}{{\lambda_{{12}}}^{2}-{\lambda_{{13}}}^{2}}}&{\dfrac {\lambda_{{
12}}}{{\lambda_{{12}}}^{2}-{\lambda_{{13}}}^{2}}}\\ \noalign{\medskip}0
&{\dfrac {\lambda_{{12}}}{{\lambda_{{12}}}^{2}-{\lambda_{{13}}}^{2}}}&-
{\dfrac {\lambda_{{13}}}{{\lambda_{{12}}}^{2}-{\lambda_{{13}}}^{2}}}
\end {array} \right),
\end{equation}
since
detA=$-{\dfrac {1}{ \left( \lambda_{{12}}-\lambda_{{13}} \right)  \left(
\lambda_{{12}}+\lambda_{{13}} \right) }}\neq 0
$
and does not depend on
$\lambda_{{23}},$
 this parameter
 can be any
value.
Furthermore,
 we also have
$\lambda_{{12}}\neq\pm\lambda_{{13}}.$
\\
First,  if
 $\lambda_{{12}}=0\neq\pm\lambda_{{13}},$ then
Jacobi structure
$(\Lambda, E)$ is classified as follows
\begin{equation}\label{IIIiii}
\Lambda^\prime=
\left( \begin {array}{ccc} 0&0&1
\\ \noalign{\medskip}0&0&0
\\ \noalign{\medskip}-1&0&0\end {array}
 \right),\qquad
E^\prime=\left( \begin {array}{c} 0\\ \noalign{\medskip} -1\\ \noalign{\medskip}1
\end {array} \right).
\end{equation}
Second,
if $\lambda_{{12}}=-\lambda_{{13}}\neq 0,$ then
 \begin{equation}
 A= \left( \begin {array}{ccc} 1&{\dfrac {-b{\lambda_{{13}}}^{2}-2\,c
\lambda_{{13}}\lambda_{{23}}+\lambda_{{23}}}{{\lambda_{{13}}}^{2}}}&b
\\ \noalign{\medskip}0&c&{\dfrac {c\lambda_{{13}}-1}{\lambda_{{13}}}}
\\ \noalign{\medskip}0&{\dfrac {c\lambda_{{13}}-1}{\lambda_{{13}}}}&c
\end {array} \right)
 \end{equation}
where
detA=${\dfrac {2\,c\lambda_{{13}}-1}{{\lambda_{{13}}}^{2}}}\neq 0,
$
hence, Jacobi structure
$(\Lambda, E)$ is classified as follows\\
\begin{equation}\label{IIIiii2}
\Lambda^{\prime \prime}=
\left( \begin {array}{ccc} 0&-1&1
\\ \noalign{\medskip}1&0&0
\\ \noalign{\medskip}-1&0&0\end {array}
 \right),\qquad
E^{\prime \prime}=\left( \begin {array}{c} 0\\ \noalign{\medskip} -2\\ \noalign{\medskip}2
\end {array} \right).
\end{equation}
Third, if $\lambda_{{12}}=\lambda_{{13}},$ then
the
Jacobi structure
$(\Lambda, E)$ is classified  the Poisson structure.
Thus, the
Jacobi structure
$(\Lambda, E)$
  is the disjoint union of equivalence classes
  $(\Lambda^{\prime}, E^{\prime}),(\Lambda^{\prime \prime}, E^{\prime \prime})$ and
    the Poisson structure.

    In this way, we have determined all of the Jacobi structures for  real  two- and three-dimensional  Lie  algebras. The results have been listed in Tables $3$ and $4$.

 The obtained  structures     can similarly be converted to   the   Jacobi   structures
$\mathbf{\Lambda^\prime}$ and $\mathbf{E^\prime}$ on Lie group.
 Thus,
we transform the Jacobi structures
$\Lambda^\prime$ and  $E^\prime$ for  the Lie algebra  to
 Jacobi structures
$\mathbf{\Lambda^\prime}$ and $\mathbf{E^\prime}$ for the Lie group using
$(\ref{basis})$ and $(\ref{basis2}).$

To compute
these Jacobi   structures, we need to determine  the vielbein $e_{a}^{\;\;\mu}$ for Lie groups, and
in order to find  the vielbein $e_{a}^{\;\;\mu}$ for Lie group,
 it is required to
calculate
 the left invariant one-forms on   the Lie group  as follows:

 \begin{equation}
 g^{-1}dg= e_{\;\;\mu}^{a}X_{a}dx^{\mu},\qquad \forall g\in \mathbf{G}
 \end{equation}
where $\lbrace X_{a} \rbrace$  are generators of the  Lie group; previously
all  of the left-invariant one forms are obtained in \textcolor{red}{ [\ref{Hemmati}, \ref{Mojaveri}]};
henceforth,  the inverse of the vielbein $e_{\;\;\mu}^{a}$ (i.e. $e_{a}^{\;\;\mu}$) for the Lie group $\mathbf{\rom{3}}$  is obtained as\footnote{Here we use the group parametrization $g= e^{x_1X_1}e^{x_2X_2}e^{x_3X_3}$ and $g= e^{x_1X_1}e^{x_2X_2}$ for three and two dimensional real Lie groups.}
\begin{equation}\label{viell}
 e_{a}^{\;\;\mu}=\left( \begin {array}{ccc} 1&0&0\\ \noalign{\medskip}-x_{{2}}-x_{{3}}
&1&0\\ \noalign{\medskip}-x_{{2}}-x_{{3}}&0&1\end {array} \right)
\end{equation}

To sum up, substituting $(\ref{IIIiii})$ and $(\ref{viell})$  in $(\ref{basis})$ and $(\ref{basis2})$ one can calculat the   Jacobi structures
$\mathbf{\Lambda^\prime}$ and  $\mathbf{E^\prime}$  on the Lie
group
$\mathbf{\rom{3}}$
as follows:
\begin{equation}
\mathbf{\Lambda^\prime}=\left( \begin {array}{ccc} 0&0&1\\ \noalign{\medskip}0&0&-x_{{2}}-x_{
{3}}\\ \noalign{\medskip}-1&x_{{2}}+x_{{3}}&0\end {array} \right),\,
\mathbf{E^\prime}=\left( \begin {array}{c} 0\\ \noalign{\medskip}-1
\\ \noalign{\medskip}1\end {array} \right).
\end{equation}
Moreover,
substituting $(\ref{IIIiii2})$ and $(\ref{viell})$  in $(\ref{basis})$ and $(\ref{basis2})$ one can calculat the  other Jacobi structures
$\mathbf{\Lambda^{\prime\prime}}$ and  $\mathbf{E^{\prime\prime}}$
on the Lie
group $\mathbf{\rom{3}}$ as follows:
\begin{equation}
\mathbf{\Lambda^{\prime\prime}}= \left( \begin {array}{ccc} 0&-1&1\\ \noalign{\medskip}1&0&-2\,x_{{2}
}-2\,x_{{3}}\\ \noalign{\medskip}-1&2\,x_{{2}}+2\,x_{{3}}&0
\end {array} \right),\,
\mathbf{E^{\prime\prime}}=\left( \begin {array}{c} 0\\ \noalign{\medskip}-2
\\ \noalign{\medskip}2\end {array} \right).
\end{equation}

\section{ Jacobi--Lie Hamiltonian systems on  real two- and three-dimensional    Lie  groups }
 In this section,  we show how  our results
 can be illustrated by some relevant examples of Jacobi--Lie Hamiltonian systems.

{\bf Example 1.}
Consider the two-dimensional real Lie group $\mathbf{A}_2$ with the  local coordinate system $\lbrace x_1, x_2 \rbrace$.

One of the
Jacobi structures on  the Lie algebra ${A}_2$ has the following forms ( see Table 3)  :
\begin{equation}\label{A222}
\Lambda=
\left( \begin {array}{cc} 0&\lambda_{{12}}
\\ \noalign{\medskip}-\lambda_{{12}}&0
\end {array}
 \right),\qquad
E=\left( \begin {array}{c} 1\\ \noalign{\medskip} 0
\end {array} \right).
\end{equation}
where
$\lambda_{12}\in \mathbb{R}-\lbrace 0 \rbrace .$
The inverse of the vielbein $e_{\;\;\mu}^{a}$ (i.e. $e_{a}^{\;\;\mu}$)  for the Lie group $\mathbf{A}_2$  is obtained as
\begin{equation}\label{A2222}
 e_{a}^{\;\;\mu}= \left( \begin {array}{cc} {e}^{x_{{2}}}&0\\ \noalign{\medskip}0&1
\end {array} \right).
\end{equation}
 Substituting $(\ref{A222})$ and $(\ref{A2222})$  in $(\ref{basis})$ and $(\ref{basis2})$ one can calculat the  Jacobi structures
$\mathbf{\Lambda}$ and  $\mathbf{E}$  on the Lie
group $\mathbf{A}_2$ as follows:
\begin{equation}\label{A2}
\mathbf{\Lambda}=
\left( \begin {array}{cc} 0&{e}^{x_{{2}}}\lambda_{{12}}
\\ \noalign{\medskip}-{e}^{x_{{2}}}\lambda_{{12}}&0\end {array}
 \right)
,\qquad\qquad
\mathbf{E}=  \left( \begin {array}{c} {e}^{x_{{2}}}\\ \noalign{\medskip}0
\end {array} \right).
\end{equation}
 It is straightforward to  show that we have:
\begin{equation*}
[[\mathbf{\Lambda, \;\Lambda}]]=2 \mathbf{E}\wedge \mathbf{\Lambda},\qquad\qquad  [[\mathbf{E,\; \Lambda}]]=0;
\end{equation*}
thus $( \mathbf{A_2,  \Lambda, E })$
is a Jacobi manifold.

Now using (\ref{A2}), (\ref{vector}),  we obtain the
Hamiltonian vector fields
\begin{equation*}
X_1^H
=
\left( -\lambda_{{12}}+x_2 \right) {e}^{ x_2}
 \partial_{x_1}
\qquad
X_2^H
=
{\frac { \left(  \left( -x_{{2}}+1 \right) {\lambda_{{12}}}^{2}+
 \left( {x_{{2}}}^{2}-x_{{2}} \right) \lambda_{{12}}+{x_{{2}}}^{2}
 \right) x_{{1}}}{ \left( x_{{2}}-\lambda_{{12}} \right) ^{2}}}
\partial_{x_1}
+{\frac {\lambda_{{12}} x_2}{x_2-\lambda_{{12}}}}
\partial_{x_2},
\end{equation*}
such that they span the   Lie algebra  ${A}_2$
with non-zero commutators
$[X_1^H,X_2^H]=X_1^H.$

  Take the
system on  $ \mathbf{A}_2$
 defined by
\begin{equation}\label{000}
\frac{d\mathfrak{\alpha}_{2}}{dt}=\sum_{i=1}^2 a_{i}(t) X_i^H(\mathfrak{\alpha}_{2}),\qquad\qquad \forall \mathfrak{\alpha}_{2}
\in  \mathbf{A}_2,
\end{equation}
for arbitrary t-dependent functions $ a_{i}(t).$

 The associated  time-dependent vector field
$X^{\mathbf{A}_2}=\sum_{i=1}^2 a_{i}(t) X_i^H$ is a Lie
   system
since $X^{\mathbf{A}_2}$
 takes values  in the Lie algebra ${A}_2.$

Meanwhile, vector fields $X_1^H$ and $ X_2^H$  are  Hamiltonian
with respect to
 $( \mathbf{{A}_2,  \Lambda, E})$
  with   Hamiltonian
 functions
 $f_1=x_2$
and
$f_2={\dfrac {{e}^{-x_2}x_2x_1}{x_2-\lambda_{{12}}}}
$, respectively. consequently,
 $( \mathbf{{A}_2,  \Lambda, E},X^{\mathbf{A}_2 })$
 is a Jacobi--Lie
  system.

Additionally, it  is easy to see that
$\lbrace f_1,f_2\rbrace_\mathbf{{\Lambda, E}}= f_1.$
 Thus,
 $X^{\mathbf{A_2}}$
 admits a Jacobi-Lie
 Hamiltonian
 system
  $( \mathbf{{A}_2,  \Lambda, E},f ).$ \\

 {\bf Example 2.}
Consider the three-dimensional real Lie group $\mathbf{ \rom{2}}$ with the  local coordinate system $\lbrace x_1, x_2,x_3 \rbrace$.
One of the
Jacobi structures on  the Lie algebra ${ \rom{2}}$ has the following forms:
\begin{equation}\label{II}
\Lambda=
\left( \begin {array}{ccc} 0&0&0
\\ \noalign{\medskip}0&0&1
\\ \noalign{\medskip}0&-1&0\end {array}
 \right),\qquad
E=\left( \begin {array}{c} 1\\ \noalign{\medskip} 0\\ \noalign{\medskip}0
\end {array} \right).
\end{equation}
 The inverse of the vielbein $e_{\;\;\mu}^{a}$ (i.e. $e_{a}^{\;\;\mu}$)  for the Lie group $\mathbf{ \rom{2}}$  is obtained as
\begin{equation}\label{vielii}
 e_{a}^{\;\;\mu}= \left( \begin {array}{ccc} 1&x_{{3}}&0\\ \noalign{\medskip}0&1&0
\\ \noalign{\medskip}0&0&1\end {array} \right).
\end{equation}
 Substituting $(\ref{II})$ and $(\ref{vielii})$  in $(\ref{basis})$ and $(\ref{basis2})$ one can calculat the  Jacobi structures
$\mathbf{\Lambda}$ and  $\mathbf{E}$  on the Lie
group
 $\mathbf{ \rom{2}}$
 as follows:
\begin{equation}\label{ex2}
 \mathbf{\Lambda}= \left( \begin {array}{ccc} 0&0&x_{{3}}\\ \noalign{\medskip}0&0&1
\\ \noalign{\medskip}-x_{{3}}&-1&0\end {array} \right)
,\qquad
 \mathbf{E}= \left( \begin {array}{c} 1\\ \noalign{\medskip} 0\\ \noalign{\medskip}0
\end {array} \right)
\end{equation}
 It is easy to prove that
$$
[[\mathbf{\Lambda,\; \Lambda}]]=2 \,\partial x_1 \wedge \partial x_2\wedge \partial x_3=2 \,\mathbf{E}\wedge \mathbf{\Lambda}, \quad  \quad [[\mathbf{E  ,\; \Lambda}]]=0;$$
so,
$( \mathbf{{\rom{2}},  \Lambda, E} )$
 is a Jacobi manifold.

Using (\ref{ex2}), (\ref{vector}),  we obtain the
Hamiltonian vector fields
\begin{equation*}
X_1^H
=
\dfrac {1}{x_2}\,\partial_{x_1}
-\dfrac {1}{{x_2}^{2}}\,\partial_{x_3},
\qquad
X_2^H
=
x_2\,\partial_{x_1}+
\partial_{x_3},
\end{equation*}
\begin{equation*}
X_3^H
=
{\frac {x_{{1}}}{2{x_{{2}}}^{2}}}\,\partial_{x_1}
-\dfrac{1}{2x_2}\,\partial_{x_2}
-{\frac {x_{{1}}}{{x_{{2}}}^{3}}}\,\partial_{x_3},
\end{equation*}
such that they span the   Lie algebra ${\rom 2}$
with non-zero commutators
$[X_2^H,X_3^H]=X_1^H.$

 Take the system on $ \mathbf{\rom 2}$ as
\begin{equation}
\frac{d\mathbb{\beta}}{dt}=\sum_{i=1}^3 a_{i}(t) X_i^H(\mathbb{\beta}),\qquad\qquad \forall \mathbb{\beta}\in  \mathbf{\rom2},
\end{equation}
for arbitrary time-dependent functions $ a_{i}(t).$

 The associated  time-dependent vector field
$X^{\mathbf{\rom2}}=\sum_{i=1}^3 a_{i}(t) X_i^H$
is a Lie system since $X^{\mathbf{\rom2}}$
 takes values  in the Lie algebra ${\rom{2}}.$
 Also,  vector fields $X_1^H,$  $ X_2^H$ and $X_3^H$   are Hamiltonian
relative to
 $( \mathbf{{\rom2},  \Lambda,  E})$
  with   Hamiltonian
 functions
 $f_1=\frac{1}{x_2},\,
f_2=x_2$ and
$f_3={\dfrac {x_{{2}}x_{{3}}+x_{{1}}}{2{x_{{2}}}^{2}}},$
 respectively.  As a result,
  $( \mathbf{{\rom2},  \Lambda, E},X^{\mathbf{\rom2} })$
 is a Jacobi--Lie  system.

Additionally,
 we can write
$\lbrace f_2,f_3\rbrace_{\mathbf{\Lambda, \, E}}= f_1.$
Therefore,
$( \mathbf{\rom2}, \mathbf{ \Lambda, E},f=\sum_{i=1}^3 a_{i}(t) f_i )$
 is a Jacobi--Lie   Hamiltonian system
 for  vector field  $X^{\mathbf{\rom2}}.$
 \\

  {\bf Example 3.}
Consider the three-dimensional real Lie group $\mathbf{ \rom{3}}$ with the  local coordinate system $\lbrace x_1, x_2,x_3 \rbrace$.  \\
One of the
Jacobi structures on  the Lie algebra ${ \rom{3}}$ has the following forms:
\begin{equation}\label{IIIIV}
\Lambda=
\left( \begin {array}{ccc}
 0&0&1
\\ \noalign{\medskip}0&0&0
\\ \noalign{\medskip}-1&0&0\end {array}
 \right),\qquad
E=\left( \begin {array}{c} 0\\ \noalign{\medskip} -1\\ \noalign{\medskip}1
\end {array} \right).
\end{equation}
 The inverse of the vielbein $e_{\;\;\mu}^{a}$ (i.e. $e_{a}^{\;\;\mu}$)  for the Lie group $\mathbf{ \rom{3}}$  is obtained as
\begin{equation}\label{vielIV1}
 e_{a}^{\;\;\mu}=\left( \begin {array}{ccc} 1&0&0\\ \noalign{\medskip}-x_{{2}}-x_{{3}}
&1&0\\ \noalign{\medskip}-x_{{2}}-x_{{3}}&0&1\end {array} \right).
\end{equation}
 Substituting $(\ref{IIIIV})$ and $(\ref{vielIV1})$  in $(\ref{basis})$ and $(\ref{basis2})$ one can calculat the  Jacobi structures
$\mathbf{\Lambda}$ and  $\mathbf{E}$  on the Lie
group as follows:
\begin{equation}\label{ex3}
 \mathbf{\Lambda}= \left( \begin {array}{ccc} 0&0&1\\ \noalign{\medskip}0&0&-x_{{2}}-x_{
{3}}\\ \noalign{\medskip}-1&x_{{2}}+x_{{3}}&0\end {array} \right)
,\qquad
 \mathbf{E}= \left( \begin {array}{c} 0\\ \noalign{\medskip} -1\\ \noalign{\medskip}1
\end {array} \right)
\end{equation}
 It is easy to check that
$$
[[\mathbf{\Lambda\; \Lambda}]]=2 \,\partial x_1 \wedge \partial x_2\wedge \partial x_3=2 \,\mathbf{E}\wedge \mathbf{\Lambda}, \quad  \quad  [[\mathbf{E  ,\; \Lambda}]]=0;$$
hence,
$( \mathbf{{\rom{3}},  \Lambda, E} )$
 is a Jacobi manifold.

Using (\ref{ex3}), (\ref{vector}),  we obtain the
Hamiltonian vector fields
\begin{equation*}
X_1^H
=
-\partial_{x_2}
+\partial_{x_3},
\qquad
X_2^H
=
-x_1\,\partial_{x_2}+
(1+x_1)\partial_{x_3},
\end{equation*}
\begin{equation*}
X_3^H
=-\partial_{x_1}
-x_1\,\partial_{x_2}+
(1+x_1)\partial_{x_3},
\end{equation*}
such that they span the   Lie algebra ${\rom 2}$
with non-zero commutators
$[X_2^H,X_3^H]=X_1^H.$
 Take the system on $ \mathbf{\rom 3}$  as
\begin{equation}
\frac{d\mathbb{\gamma}}{dt}=\sum_{i=1}^3 a_{i}(t) X_i^H(\mathbb{\gamma}),\qquad\qquad \mathbb{\gamma}\in  \mathbf{\rom3},
\end{equation}
for arbitrary t-dependent functions $ a_{i}(t).$

 The associated  t-dependent vector field
$X^{\mathbf{\rom3}}=\sum_{i=1}^3 a_{i}(t) X_i^H$
 is a Lie system
since $X^{\mathbf{\rom3}}$
takes values  in the Lie algebra ${\rom{2}}.$

 Meanwhile,   vector fields $X_1^H$, $ X_2^H$ and $X_3^H$   are Hamiltonian
relative to
 $( \mathbf{{\rom3},  \Lambda,  E})$
  with   Hamiltonian
 functions
 $f_1=1,$
$f_2=x_1$ and $f_3= x_1+x_2+x_3,
$ respectively. Thus,
  $( \mathbf{{\rom3},  \Lambda, E},X^{\mathbf{\rom3} })$
 is a Jacobi--Lie  system.

In addition,
 we can write
$\lbrace f_2,f_3\rbrace_{\mathbf{\Lambda, \, E}}= f_1.
$
Then,
$( \mathbf{{\rom3},  \Lambda, E},f=\sum_{i=1}^3 a_{i}(t) f_i )$
 is a Jacobi--Lie   Hamiltonian system
 for  vector field $X^{\mathbf{\rom3}}$
 .\\
Another  equivalence class of
Jacobi structures on  the Lie algebra ${ \rom{3}}$ has the following forms:
\begin{equation}\label{III2}
\Lambda^\prime=
\left( \begin {array}{ccc}
 0&-1&1
\\ \noalign{\medskip}1&0&0
\\ \noalign{\medskip}-1&0&0\end {array}
 \right),\qquad
E^\prime=\left( \begin {array}{c} 0\\ \noalign{\medskip} -2\\ \noalign{\medskip}2
\end {array} \right).
\end{equation}
 Substituting $(\ref{III2})$ and $(\ref{vector})$  in $(\ref{basis})$ and $(\ref{basis2})$ one can calculat the  Jacobi structures
$\mathbf{\Lambda^\prime}$ and  $\mathbf{E}^\prime$  on the Lie
group as follows:
\begin{equation}\label{ex3tow}
 \mathbf{\Lambda}^\prime=  \left( \begin {array}{ccc} 0&-1&1\\ \noalign{\medskip}1&0&-2\,x_{{2}
}-2\,x_{{3}}\\ \noalign{\medskip}-1&2\,x_{{2}}+2\,x_{{3}}&0
\end {array} \right)
,\qquad
 \mathbf{E}^\prime= \left( \begin {array}{c} 0\\ \noalign{\medskip} -2\\ \noalign{\medskip}2
\end {array} \right)
\end{equation}
 It is easy to check that
$$
[[\mathbf{\Lambda^\prime,\; \Lambda^\prime}]]=0=2 \,\mathbf{E}^\prime\wedge \mathbf{\Lambda}^\prime, \quad  \quad  [[\mathbf{E ^\prime ,\; \Lambda^\prime}]]=0;$$
Hence,
$( \mathbf{{\rom{3}},  \Lambda^\prime, E^\prime} )$
 is a Jacobi manifold.\\
 Using (\ref{ex3tow}) and (\ref{vector}),
we see that
 the
 vector fields
$X_1, X_2, X_3$
cannot form a basis for
  the  Lie
group of
$ \mathbf{\rom3}.$
 So we
 cannot
discuss the
Lie system.\\
  {\bf Example 4.}
Consider the three-dimensional real Lie group $\mathbf{ \rom{4}}$ with the local coordinate system $\lbrace x_1, x_2,x_3 \rbrace$.
One of the
Jacobi structures on  the Lie algebra ${ \rom{4}}$ has the following forms:

\begin{equation}\label{IVtow}
\Lambda^\prime=
\left( \begin {array}{ccc} 0&1&0\\ \noalign{\medskip}-1&0&0
\\ \noalign{\medskip}0&0&0\end {array} \right)
,\qquad
E^\prime=\left( \begin {array}{c} 0\\ \noalign{\medskip} 1\\ \noalign{\medskip}1
\end {array} \right).
\end{equation}
 The inverse of the vielbein $e_{\;\;\mu}^{a}$ (i.e. $e_{a}^{\;\;\mu}$)  for the Lie group $\mathbf{ \rom{4}}$  is obtained as
\begin{equation}\label{vieliv4tow}
 e_{a}^{\;\;\mu}=\left( \begin {array}{ccc} 1&0&0\\ \noalign{\medskip}-x_{{2}}&1&0
\\ \noalign{\medskip}x_{{2}}-x_{{3}}&0&1\end {array} \right).
\end{equation}
 Substituting (\ref{IVtow}) and (\ref{vieliv4tow})  in $(\ref{basis})$ and $(\ref{basis2})$ one can calculat the  Jacobi structures
$\mathbf{\Lambda}^\prime$ and  $\mathbf{E}^\prime$  on the Lie
group as follows:
\begin{equation}\label{ex4two}
 \mathbf{\Lambda}^\prime=  \left( \begin {array}{ccc} 0&1&0\\ \noalign{\medskip}-1&0&-x_{{2}}+x
_{{3}}\\ \noalign{\medskip}0&x_{{2}}-x_{{3}}&0\end {array} \right)
,\qquad
 \mathbf{E}^\prime= \left( \begin {array}{c} 0\\ \noalign{\medskip} 1\\ \noalign{\medskip}1
\end {array} \right)
\end{equation}
 It is easy to check that
$$
[[\mathbf{\Lambda^\prime,\; \Lambda^\prime}]]=2 \,\partial x_1 \wedge \partial x_2\wedge \partial x_3=2 \,\mathbf{E}^\prime\wedge \mathbf{\Lambda}^\prime, \quad  \quad  [[\mathbf{E^\prime  ,\; \Lambda^\prime}]]=0;$$
Hence,
$( \mathbf{{\rom{4}},  \Lambda^\prime, E^\prime} )$
 is a Jacobi manifold.
Using (\ref{ex4two}), (\ref{vector}),  we obtain the
Hamiltonian vector fields
$
X_1^H
=
\partial_{x_2}
+\partial_{x_3},
$
\begin{equation*}
X_2^H
=
(-{\frac {{{\rm e}^{-x_{{3}}}}}{x_{{2}}-x_{{3}}}})\,\partial_{x_1}+
(- \left( x_{{3}}+\ln  \left( -x_{{2}}+x_{{3}} \right)  \right)
 \left( -1+x_{{2}}-x_{{3}} \right) {{\rm e}^{-x_{{3}}}}
)\partial_{x_2}+
({{\rm e}^{-x_{{3}}}} \left( x_{{3}}+\ln  \left( -x_{{2}}+x_{{3}}
 \right)  \right)
)\partial_{x_3}
\end{equation*}
and
\begin{equation*}
X_3^H
=
-{{\rm e}^{-x_{{3}}}} \left( -1+x_{{2}}-x_{{3}} \right))\,\partial_{x_2}+
({{\rm e}^{-x_{{3}}}}
)\partial_{x_3},
\end{equation*}
such that they span the   Lie algebra $\rom{ 4}$
with non-zero commutators
$[X_1^H,X_2^H]=-X_2^H+X_3^H, [X_1^H,X_3^H]=-X_3^H.$
Take the system on $ \mathbf{\rom 4}$  as
\begin{equation}
\frac{d\mathbb{\delta}}{dt}=\sum_{i=1}^3 a_{i}(t) X_i^H(\mathbb{\delta}),\qquad\qquad \forall \mathbb{\delta}\in  \mathbf{\rom 4},
\end{equation}
for arbitrary time-dependent functions $ a_{i}(t).$
 The associated  time-dependent vector field
$X^{\mathbf{\rom4}}=\sum_{i=1}^3 a_{i}(t) X_i^H$
 is a Lie system
since $X^{\mathbf{\rom4}}$
takes values  in the Lie algebra ${\rom{4}}.$ \\
  Moreover, vector fields $X_1^H, \, X_2^H$ and $X_3^H$   are Hamiltonian
relative to
 $( \mathbf{{\rom4},  \Lambda^\prime, E^\prime} )$
  with   Hamiltonian
 functions
 $f_1=1,  \,f_2=
\left( \ln  \left( -y+z \right) +1+z \right) {{\rm e}^{-z}}
$ and $f_3=e^{-z},$
 respectively. Consequently,
  $( \mathbf{\rom{4},  \Lambda^\prime, E^\prime}, X^{\mathbf{\rom{4} }})$
 is a Jacobi--Lie system.
 Additionally,
  we have
  $\lbrace f_1,f_2\rbrace_\mathbf{\Lambda^\prime, \, E^\prime}=- f_2+f_3,\quad
\lbrace f_1,f_3\rbrace_\mathbf{\Lambda^\prime, \, E^\prime}=- f_3.$\\
So,
$( \mathbf{{\rom4},  \Lambda^\prime, E^\prime},f=\sum_{i=1}^3 a_{i}(t)f_i  )$
 is a Jacobi--Lie   Hamiltonian system
 for  vector field  $X^{ \mathbf{\rom{4}}}.$ \\
 Another  equivalence class of
Jacobi structures on  the Lie algebra ${ \rom{4}}$ has the following forms:
\begin{equation}\label{IVIV}
\Lambda=
\left( \begin {array}{ccc} 0&0&1\\ \noalign{\medskip}0&0&0
\\ \noalign{\medskip}-1&0&0\end {array} \right)
,\qquad
E=\left( \begin {array}{c} 0\\ \noalign{\medskip} 0\\ \noalign{\medskip}1
\end {array} \right).
\end{equation}
 Substituting $(\ref{vieliv4tow})$ and $(\ref{IVIV})$  in $(\ref{basis})$ and $(\ref{basis2})$ one can calculat the  Jacobi structures
$\mathbf{\Lambda}$ and  $\mathbf{E}$  on the Lie
group as follows:
\begin{equation}\label{ex4one}
 \mathbf{\Lambda}= \left( \begin {array}{ccc} 0&0&1\\ \noalign{\medskip}0&0&-x_{{2}}
\\ \noalign{\medskip}-1&x_{{2}}&0\end {array} \right)
,\qquad
 \mathbf{E}= \left( \begin {array}{c} 0\\ \noalign{\medskip} 0\\ \noalign{\medskip}1
\end {array} \right)
\end{equation}
 It is easy to check that
$$
[[\mathbf{\Lambda\; \Lambda}]]=0=2 \,\mathbf{E}\wedge \mathbf{\Lambda}, \quad  \quad  [[\mathbf{E  ,\; \Lambda}]]=0;$$
hence,
$( \mathbf{{\rom{4}},  \Lambda, E} )$
 is a Jacobi manifold.
 Using (\ref{ex4one}) and (\ref{vector}),
we see that
 the
 vector fields
$X_1, X_2, X_3$
cannot form a basis for
  the  Lie
group of
$ \mathbf{\rom{4}}.$
 So we
 cannot
discuss the
Lie system. \\

 {\bf Example 5.}
Consider the three-dimensional real Lie group $\mathbf{ \rom{6}_0}$ with the  local coordinate system $\lbrace x_1, x_2,x_3 \rbrace$.
One of the
Jacobi structures on  the Lie algebra $ \rom{6}_0$ has the following forms:
\begin{equation}\label{VI0}
\Lambda=
\left( \begin {array}{ccc} 0&0&0
\\ \noalign{\medskip}0&0&1
\\ \noalign{\medskip}0&-1&0\end {array}
 \right),\qquad
E=\left( \begin {array}{c} 1\\ \noalign{\medskip} 0\\ \noalign{\medskip}0
\end {array} \right).
\end{equation}
 The inverse of the vielbein $e_{\;\;\mu}^{a}$ (i.e. $e_{a}^{\;\;\mu}$)  for the Lie group $\mathbf{ \rom{6}_0}$  is obtained as
\begin{equation}\label{vielvi0}
 e_{a}^{\;\;\mu}= \left( \begin {array}{ccc} \cosh \left( x_{{3}} \right) &\sinh
 \left( x_{{3}} \right) &0\\ \noalign{\medskip}\sinh \left( x_{{3}}
 \right) &\cosh \left( x_{{3}} \right) &0\\ \noalign{\medskip}0&0&1
\end {array} \right).
\end{equation}
 Substituting $(\ref{VI0})$ and $(\ref{vielvi0})$  in $(\ref{basis})$ and $(\ref{basis2})$ one can calculat the  Jacobi structures
$\mathbf{\Lambda}$ and  $\mathbf{E}$  on the Lie
group as follows:
\begin{equation}\label{ex5}
 \mathbf{\Lambda}=\left( \begin {array}{ccc} 0&0&\sinh \left( x_{{3}} \right)
\\ \noalign{\medskip}0&0&\cosh \left( x_{{3}} \right)
\\ \noalign{\medskip}-\sinh \left( x_{{3}} \right) &-\cosh \left( x_{{
3}} \right) &0\end {array} \right)
,\qquad
 \mathbf{E}=\left( \begin {array}{c} \cosh \left( x_{{3}} \right)
\\ \noalign{\medskip}\sinh \left( x_{{3}} \right)
\\ \noalign{\medskip}0\end {array} \right)
\end{equation}
 It is easy to prove that
$$
[[\mathbf{\Lambda\; \Lambda}]]=2 \,\partial x_1 \wedge \partial x_2\wedge \partial x_3=2 \,\mathbf{E}\wedge \mathbf{\Lambda}, \quad  \quad  [[\mathbf{E  ,\; \Lambda}]]=0;$$
so,
$( \mathbf{{\rom{2}},  \Lambda, E} )$
 is a Jacobi manifold.

Using (\ref{ex5}), (\ref{vector}),  we obtain the
Hamiltonian vector fields
$
X_1^H
=
\cosh \left( x_{{3}} \right)\,\partial_{x_1}
+\sinh \left( x_{{3}} \right)\,\partial_{x_2},
$
\begin{equation*}
X_2^H
=
(-\sinh \left( x_{{3}} \right) +x_{{3}}\cosh \left( x_{{3}} \right) )\,\partial_{x_1}+
(-\cosh \left( x_{{3}} \right) +x_{{3}}\sinh \left( x_{{3}} \right))\partial_{x_2},
\end{equation*}
and
$
X_3^H
=
-x_{2}\,\partial_{x_1}
-x_{1}\,\partial_{x_2}
-\,\partial_{x_3},
$
such that they span the   Lie algebra ${\rom 2}$
with non-zero commutators
$[X_2^H,X_3^H]=X_1^H.$
Take The system on $ \mathbf{\rom 6_0}$  as
\begin{equation}
\frac{d\mathbb{\zeta}}{dt}=\sum_{i=1}^3 a_{i}(t) X_i^H(\mathbb{\zeta}),\qquad\qquad \forall \mathbb{\zeta}\in  \mathbf{\rom 6_0},
\end{equation}
for arbitrary t-dependent functions $ a_{i}(t).$
 The associated  time-dependent vector field
$X^{\mathbf{\rom6}_0}=\sum_{i=1}^3 a_{i}(t) X_i^H$
is a Lie system
since $X^{\mathbf{\rom6_0}}$
 takes values  in the Lie algebra ${\rom2}.$

 In addition,  vector fields $X_1^H, X_2^H$ and $X_3^H$   are Hamiltonian
relative to
 $( \mathbf{\rom6_0,  \Lambda, E} )$
  with   Hamiltonian
 functions
 $f_1=1, f_2=x_3$
and
$ f_3=x_{{1}}\sinh \left( x_{{3}} \right) -x_{{2}}\cosh \left( x_{{3}}
 \right),$
 respectively. Thus,
  $( \mathbf{\rom6_0,  \Lambda, E}, X^{\mathbf{\rom6_0 }})$
 is a Jacobi--Lie system.

 Additionally,
  the
functions
$f_1,f_2$
and
$f_3$ satisfy
$
\lbrace f_2,f_3\rbrace_{\mathbf{\Lambda_2, \, E_2 }}= f_1.$
Then,
$( \mathbf{\rom6_0,  \Lambda, E},f=\sum_{i=1}^3 a_{i}(t)  f_i)$
 is a Jacobi--Lie   Hamiltonian system
 for vector field  $X^{ \mathbf{\rom6_0}}$
 .\\

 {\bf Example 6.}
Consider the three-dimensional real Lie group $\mathbf{ \rom{7}_0}$ with the  local coordinate system $\lbrace x_1, x_2,x_3 \rbrace$.
One of the
Jacobi structures on  the Lie algebra $ \rom{7}_0$ has the following forms:
\begin{equation}\label{VII0}
\Lambda=
\left( \begin {array}{ccc}
 0&0&0
\\ \noalign{\medskip}0&0&1
\\ \noalign{\medskip}0&-1&0\end {array}
 \right),\qquad
E=\left( \begin {array}{c} 1\\ \noalign{\medskip} 0\\ \noalign{\medskip}0
\end {array} \right).
\end{equation}
 The inverse of the vielbein $e_{\;\;\mu}^{a}$ (i.e. $e_{a}^{\;\;\mu}$)  for the Lie group $\mathbf{ \rom{7}_0}$  is obtained as
\begin{equation}\label{vielvii0}
 e_{a}^{\;\;\mu}=
 \left( \begin {array}{ccc} \cos \left( x_{{3}} \right) &\sin \left( x
_{{3}} \right) &0\\ \noalign{\medskip}-\sin \left( x_{{3}} \right) &
\cos \left( x_{{3}} \right) &0\\ \noalign{\medskip}0&0&1\end {array}
 \right).
\end{equation}
 Substituting $(\ref{VII0})$ and $(\ref{vielvii0})$  in $(\ref{basis})$ and $(\ref{basis2})$ one can calculat the  Jacobi structures
$\mathbf{\Lambda}$ and  $\mathbf{E}$  on the Lie
group as follows:
\begin{equation}\label{ex6}
 \mathbf{\Lambda}=\left( \begin {array}{ccc} 0&0&\sin \left( x_{{3}} \right)
\\ \noalign{\medskip}0&0&\cos \left( x_{{3}} \right)
\\ \noalign{\medskip}-\sin \left( x_{{3}} \right) &-\cos \left( x_{{3}
} \right) &0\end {array} \right)
,\qquad
 \mathbf{E}=\left( \begin {array}{c} \cos \left( x_{{3}} \right)
\\ \noalign{\medskip}-\sin \left( x_{{3}} \right)
\\ \noalign{\medskip}0\end {array} \right)
\end{equation}
 It is easy to prove that
$$
[[\mathbf{\Lambda\; \Lambda}]]=2 \,\partial x_1 \wedge \partial x_2\wedge \partial x_3=2 \,\mathbf{E}\wedge \mathbf{\Lambda}, \quad  \quad  [[\mathbf{E  ,\; \Lambda}]]=0;$$
so,
$( \mathbf{{\rom{2}},  \Lambda, E} )$
 is a Jacobi manifold.

Using (\ref{ex6}), (\ref{vector}),  we obtain the
Hamiltonian vector fields
\begin{equation*}
X_1^H
=
\cos \left( x_{{3}} \right)\,\partial_{x_1}
-\sin \left( x_{{3}} \right)\,\partial_{x_2},
\qquad
X_2^H
=
(-\sin \left( x_{{3}} \right) +x_{{3}}\cos \left( x_{{3}} \right) )\,\partial_{x_1}+
(-\cos \left( x_{{3}} \right) -x_{{3}}\sin \left( x_{{3}} \right) )\partial_{x_2},
\end{equation*}
\begin{equation*}
X_3^H=
(\left( -2\,x_{{1}}x_{{3}}+x_{{2}} \right)  \left( \cos \left( x_{{3}}
 \right)  \right) ^{2}+ \left( 2\,x_{{2}}x_{{3}}+x_{{1}} \right) \cos
 \left( x_{{3}} \right) \sin \left( x_{{3}} \right) -x_{{2}}{x_{{3}}}^
{2}+x_{{1}}x_{{3}}-2\,x_{{2}}
)\,\partial_{x_1}
+
\end{equation*}
\begin{equation*}
(\left( 2\,x_{{2}}x_{{3}}+x_{{1}} \right)  \left( \cos \left( x_{{3}}
 \right)  \right) ^{2}+2\,\sin \left( x_{{3}} \right)  \left( x_{{1}}x
_{{3}}-\frac{1}{2}\,x_{{2}} \right) \cos \left( x_{{3}} \right) +x_{{1}}{x_{{3
}}}^{2}-x_{{2}}x_{{3}}+x_{{1}}
)\,\partial_{x_2}
+(-{x_{{3}}}^{2}-1)\partial_{x_3},
\end{equation*}

such that they span the   Lie algebra ${\rom{7}}_0$ with non-zero commutators
$[X_1^H,X_3^H]=-X_2^H, [X_2^H,X_3^H]=X_1^H.$

 Take The system on $ \mathbf{\rom 7_0}$ as

\begin{equation}
\frac{d\mathbb{\eta}}{dt}=\sum_{i=1}^3 a_{i}(t) X_i^H(\mathbb{\eta}),\qquad\qquad \forall \mathbb{\eta}\in  \mathbf{\rom7_0},
\end{equation}
for arbitrary non-autonomous functions $ a_{i}(t).$

The associated non-autonomous vector field
$X^{\mathbf{\rom7_0}}=\sum_{i=1}^3 a_{i}(t) X_i^H$
 is a Lie system since  $X^{\mathbf{\rom7_0}}$
takes values  in the Lie algebra  ${\rom{7}}_0.$
 Moreover, vector fields
  $X_1^H, X_2^H$ and $X_3^H$   are Hamiltonian
relative to
 $( \mathbf{\rom7_0,  \Lambda, E} )$
  with   Hamiltonian
 functions
 $f_1=1,\,
f_2=x_3$ and
$f_3= \left( -x_{{2}}{x_{{3}}}^{2}-x_{{1}}x_{{3}}-x_{{2}} \right) \cos
 \left( x_{{3}} \right) -\sin \left( x_{{3}} \right)  \left( x_{{1}}{x
_{{3}}}^{2}-x_{{2}}x_{{3}}+x_{{1}} \right)
$
 respectively.
Then,
  $( \mathbf{\rom7_0,  \Lambda, E}, X^{\mathbf{\rom7_0 }})$
 is a Jacobi--Lie system.

 Additionally, we observe
 that
$\lbrace f_1,f_3\rbrace_{\mathbf{\Lambda, \, E }}=- f_2$ and $
\lbrace f_2,f_3\rbrace_{\mathbf{\Lambda, \, E }}= f_1.$\\
Therefore,
$( \mathbf{\rom7_0,  \Lambda, E},f=\sum_{i=1}^3 a_{i}(t) f_i )$
 is a Jacobi--Lie   Hamiltonian system
 for  vector field $X^{ \mathbf{\rom7_0}}$
 .\\

\newpage
\section*{Appendix 1: Real Two- and Three-Dimensional Lie Algebras}

{\bf Table 1}:
Real two-dimensional Lie
algebras.\\
    \begin{tabular}{|l| l| }
    \hline

    \hline
   Lie algebra\qquad\qquad\qquad &    Commutation relations\\
\hline
$A_{1}$ & {\footnotesize $[X_{i},X_{j}]=0$}\\
\hline
$A_{2}$& {\footnotesize$ [X_{1},X_{2}]=X_{1}$}\\
\hline
\end{tabular}

\vspace*{0.5cm}
{\bf Table 2}:
Real three-dimensional Lie algebras.\\
 \begin{tabular}{|l| l|  l| }
 \hline

 \hline
   Lie algebra &    Commutation relations &  Comments\\
\hline
${ \rom{1}}$& {\footnotesize $[X_{i},X_{j}]=0$} &\\
\hline
${ \rom{2}}$& {\footnotesize $[X_{2},X_{3}]=X_{1}$} & \\
\hline
${ \rom{3}}$& {\footnotesize$[X_{1},X_{3}]=-(X_{2}+X_{3})$} $,$ {\footnotesize $[X_{1},X_{2}]=-(X_{2}+X_{3})$} & \\
\hline
${ \rom{4}}$& {\footnotesize$[X_{1},X_{3}]=-X_{3}$}  $,$ {\footnotesize $[X_{1},X_{2}]=-(X_{2}-X_{3})$}&  \\
\hline
${ \rom{5}}$& {\footnotesize$[X_{1},X_{3}]=-X_{3}$} $,$ {\footnotesize $[X_{1},X_{2}]=-X_{2}$} &  \\
\hline
${ \rom{6}_0}$& {\footnotesize$[X_{2},X_{3}]=X_{1}$} $,$ {\footnotesize $[X_{1},X_{3}]=X_{2}$}& \\
\hline
 ${ \rom{6}_a}$& {\footnotesize$[X_{1},X_{3}]=-(X_{2}+aX_{3})$} $,$ {\footnotesize $[X_{1},X_{2}]=-(aX_{2}+X_{3})$} &  {\footnotesize $ a\in\mathbb{R}  -\{1\},\;\;a>0$ }\\
\hline
${ \rom{7}_0}$& {\footnotesize$[X_{2},X_{3}]=X_{1}$ $,$ {\footnotesize $[X_{1},X_{3}]=-X_{2}$}} & \\
\hline
 ${ \rom{7}_a}$& {\footnotesize$[X_{1},X_{3}]=-(X_{2}+aX_{3})$} $,$ {\footnotesize $[X_{1},X_{2}]=-(aX_{2}-X_{3})$} &  {\footnotesize $a\in\mathbb{R}, \;\;a>0 $} \\
\hline
${ \rom{8}}$& {\footnotesize$[X_{2},X_{3}]=X_{1}$} $,$ {\footnotesize$[X_{1},X_{3}]=-X_{2}$} $,$ {\footnotesize $[X_{1},X_{2}]=-X_{3}$}  & \\
\hline
${ \rom{9}}$& {\footnotesize$[X_{2},X_{3}]=X_{1}$} $,$ {\footnotesize$[X_{1},X_{3}]=-X_{2}$} $,$ {\footnotesize $[X_{1},X_{2}]=X_{3}$} & \\
\hline
\hline
\end{tabular}

\section*{Appendix 2: Jacobi Structures  on Two- and Three-Dimensional Lie Algebras and  Equivalence Classes}
{\bf Table 3}:
Jacobi structures  on two-dimensional  Lie algebras and  Equivalence classes.\\
    \begin{tabular}{|l| l|  l| }
    \hline

    \hline
 Jacobi structure  on  Lie algebra $A_1$ &  Equivalence classes &Comments\qquad\qquad\qquad\qquad
 \\
\hline

$\Lambda=\lambda_{12}\partial _{x_1}\wedge \partial_{ x_2}$
&
$\Lambda=\partial _{x_1}\wedge \partial_{ x_2}$
&
\\
 ${E}= -e_{1}\partial _{x_1} -e_{2}\partial _{x_2}$
&
${E}= -\partial _{x_1}$
&\\
\hline

 Jacobi structures  on  Lie algebra $A_2$ &&
 \\
\hline

${\Lambda}_1=\lambda_{12}\partial _{x_1}\wedge \partial_{ x_2}$
&
$\Lambda= \lambda_{12}\partial _{x_1}\wedge \partial_{ x_2}$
&
$\lambda_{12} \in \mathbb{R}-\lbrace 0\rbrace$
 \\
 $E_1=e_{1}\partial _{x_1}
$
 &
$E=\partial _{x_1}
$
&
 \\
 \hline

 ${\Lambda}_2=0
$
&
 $\Lambda=0
$&
 \\
${E}_2= -e_{1}\partial _{x_1} -e_{2}\partial _{x_2}$
&
${E}= -\partial _{x_1} -e_{2}\partial _{x_2}$
&$e_{2} \in \mathbb{R}$
  \\
\hline
        \end{tabular}

       \vspace*{0.5cm}

{\bf Table 4}:
Jacobi structures  on Bianchi real three dimensional  Lie algebras and   Equivalence classes.\\
    \begin{tabular}{| l|l |l |   }
      \hline

      \hline
{\footnotesize Jacobi structures    on  Lie algebra \rom{1}} &{\footnotesize Equivalence classes }&
{\footnotesize Comments}\\
 \hline

 $ {\Lambda}_1 =\dfrac {-e_{{1}}\lambda_{{23}}+e_{{2}}
\lambda_{{13}}}{e_{{3}}}\partial _{x_1}\wedge \partial_{ x_2}+
\lambda_{{13}}\partial _{x_1}\wedge \partial_{ x_3}+\lambda_{{23
}}\partial _{x_2}\wedge \partial_{ x_3}
 $
&
 $\Lambda=\partial _{x_1}\wedge \partial_{ x_3}+
 \partial _{x_2}\wedge \partial_{ x_3}
$
&
\\

 ${E}_1=- e_{{1}}\partial _{x_1}-e_{{2}}\partial_{ x_2}
-e_{{3}}\partial _{x_3}
$
&
 $E=-\partial _{x_3}
 $
 &
 \\
&
&\\
  \hline

${\Lambda}_2=
\lambda_{{12}}\partial _{x_1}\wedge \partial_{ x_2}+\frac {e_{{1}}\lambda_{{
23}}}{e_{{2}}}\partial _{x_1}\wedge \partial_{ x_3}+\lambda_{{23}}\partial _{x_2}\wedge \partial_{ x_3}
$
&
$\Lambda=
\partial _{x_1}\wedge \partial_{ x_2}$
&

\\
 ${E}_2=- e_{{1}}\partial _{x_1}-e_{{2}}\partial_{ x_2}
$
&
$E=-\partial_{ x_2}$
&
\\
 \hline

${\Lambda}_3=
\lambda_{{12}}\partial _{x_1}\wedge \partial_{ x_2}+\lambda_{{13}}\partial _{x_1}\wedge \partial_{ x_3}
$
&
$\Lambda
=\partial _{x_1}\wedge \partial_{ x_2}+
 \partial _{x_1}\wedge \partial_{ x_3}$
&

 \\
 ${E}_3= - e_{{1}}\partial _{x_1}
$
&
$E=-\partial _{x_1}$
&
\\
\hline

      \hline
{\footnotesize Jacobi structures  on  Lie algebra \rom{2}}
 &{\footnotesize  Equivalence classes  }
 &
\\
 \hline
 $ {\Lambda}_1 =\lambda_{{12}}\partial _{x_1}\wedge \partial_{ x_2}+\lambda_{{13}}\partial _{x_1}\wedge \partial_{ x_3}$
&
$\Lambda= \partial _{x_1}\wedge \partial_{ x_2}$
&
\\

${E}_1=- e_{{1}}\partial _{x_1}-e_{{2}}\partial_{ x_2}
-\frac {e_{{2}}\lambda_{{
13}}}{\lambda_{12}}\partial _{x_3} $
&
$E= - \partial _{x_1}-\partial_{ x_2}
$
&
 \\ \hline

${\Lambda}_2=
\lambda_{{12}}\partial _{x_1}\wedge \partial_{ x_2}+\lambda_{{13}}\partial _{x_1}\wedge \partial_{ x_3}+
\lambda_{{23}}\partial _{x_2}\wedge \partial_{ x_3}
$
&
$\Lambda= \partial _{x_2}\wedge \partial_{ x_3}$
&
 \\

 ${E}_2=\lambda_{23} \partial _{x_1}$
&
$E= \partial _{x_1}$
&
\\  \hline
     \end{tabular}

   \newpage
{\bf Table 4}:
(Continued.)\\
    \begin{tabular}{|  l| l| l|   }
      \hline

      \hline
{\footnotesize Jacobi structures  on  Lie algebra ${\rom{3}}$} &{\footnotesize Equivalence classes }&
{\footnotesize Comments}
\\
 \hline
$ {\Lambda}_1 =\lambda_{{12}}\partial _{x_1}\wedge \partial_{ x_2}-\lambda_{{12}}\partial _{x_1}\wedge \partial_{ x_3}+
\lambda_{{23}}\partial _{x_2}\wedge \partial_{ x_3} $
&
$ \Lambda=\partial _{x_2}\wedge \partial_{ x_3}
 $
 &
\\
${E}_1=\frac {\lambda_{12} \left( e_{{3}}+e_{{2}} \right) }{
\lambda_{23}}\partial _{x_1}-e_{{2}}\partial _{x_2}-e_{{3}}\partial _{x_1}
 $
&
$E= -\partial _{x_3} $
& \\ \hline

${\Lambda}_2=
\lambda_{{12}}\partial _{x_1}\wedge \partial_{ x_2}+\lambda_{{13}}\partial _{x_1}\wedge \partial_{ x_3}+
\lambda_{{23}}\partial _{x_2}\wedge \partial_{ x_3}$
&
$\Lambda=\partial _{x_1}\wedge \partial_{ x_3}
$
&
$  \begin {array}{c} \lambda_{12}\neq \lambda_{13}\\ \noalign{\medskip}\lambda_{12}\neq -\lambda_{13}
\end {array}$
 \\
${E}_2=(-\lambda_{{13}}+
\lambda_{{12}})\partial _{x_2}+(\lambda_{{13}}-\lambda_{{12}})\partial _{x_3}
$
&
$E=-\partial _{x_2}+\partial _{x_3} $
&
\\

&

&
\\

&
$\Lambda^\prime=-\partial _{x_1}\wedge \partial_{ x_2}+\partial _{x_1}\wedge \partial_{ x_3}
$
&
$  \begin {array}{c} \lambda_{12}= -\lambda_{13}
\end {array}$
 \\
&
$E^\prime= -2\partial _{x_2}+2\partial _{x_3} $
&
\\

\hline

$ {\Lambda}_3 =\lambda_{{12}}\partial _{x_1}\wedge \partial_{ x_2}+\lambda_{{12}}\partial _{x_1}\wedge \partial_{ x_3}+
\lambda_{{23}}\partial _{x_2}\wedge \partial_{ x_3} $
&
$ \Lambda=\partial _{x_2}\wedge \partial_{ x_3}
 $
&
$
 \begin {array}{c} \lambda_{12}=0\\ \noalign{\medskip}e_{{2}}\neq 0
\end {array}
$
 \\
${E}_3=-e_{{2}}\partial _{x_2}-e_{{2}}\partial _{x_3}
 $
&
$E= -\partial _{x_2}  -\partial _{x_3}$
& \\

&

&
\\
&
$\Lambda^\prime=\partial _{x_1}\wedge \partial_{ x_2}+\partial _{x_1}\wedge \partial_{ x_3}+\partial _{x_2}\wedge \partial_{ x_3}
$
&
$\lambda_{{12}}=e_{{2}}\neq 0$
\\
&
$
E^\prime= -\partial _{x_2}  -\partial _{x_3}
$
&
\\\hline

\hline
{\footnotesize Jacobi structures  on  Lie algebra ${\rom{4}}$ } &{\footnotesize   Equivalence classes}
&
\\
 \hline
 $ {\Lambda}_1 =
0
 $
&
 ${ \Lambda} =
0
$
&
\\
 ${E}_1=- e_{{1}}\partial _{x_1}-e_{{2}}\partial_{ x_2}
-e_{{3}}\partial _{x_3}
$
&
${E}=
- e_{{1}}\partial _{x_1}
 $
 &$e_{1} \in \mathbb{R}-\lbrace 0\rbrace$
 \\ \hline

${\Lambda}_2=
\lambda_{{12}}\partial _{x_1}\wedge \partial_{ x_2}+\lambda_{{13}}\partial _{x_1}\wedge \partial_{ x_3}+
\lambda_{{23}}\partial _{x_2}\wedge \partial_{ x_3}
$
&
$\Lambda=\partial _{x_1}\wedge \partial_{ x_3}
$
&
$\begin {array}{c}\lambda_{12} = 0\\ \noalign{\medskip}\lambda_{13} \neq 0\end {array}$
 \\

 ${E}_2= \lambda_
{{12}}\partial _{x_2}+(\lambda_{{12}}+\lambda_{{13}})\partial _{x_3}
$
&
${E}=
\partial _{x_3}
 $
 &

\\
&
 &

\\

&
$\Lambda^\prime=\partial _{x_1}\wedge \partial_{ x_2}
$
&
 $\begin {array}{c}\lambda_{12} \neq 0\end {array}$
\\
&
 $E^\prime=
\partial _{x_2}+\partial _{x_3} $
 &

\\

 \hline

${\Lambda}_3=
\lambda_{{13}}\partial _{x_1}\wedge \partial_{ x_3}+
\lambda_{{23}}\partial _{x_2}\wedge \partial_{ x_3}
$
&
$\Lambda=\lambda_{{13}}\partial _{x_1}\wedge \partial_{ x_3}
$
&
$\lambda_{13} \in \mathbb{R}-\lbrace 0\rbrace$
 \\
 ${E}_3= -e_{{3}}\partial _{x_3}
$
&$E=-\partial _{x_3}
$
&\\  \hline

${\Lambda}_4=
\lambda_{{23}}\partial _{x_2}\wedge \partial_{ x_3}
$
&
$
\Lambda= \lambda_{{23}}\partial _{x_2}\wedge \partial_{ x_3}
$
&
$\lambda_{23} \in \mathbb{R}-\lbrace 0\rbrace$
 \\
${E}_4=  -e_{2}\partial _{x_2}
-e_{3}\partial _{x_3}
$
&$E=
 -\partial _{x_2}-\partial _{x_3}
$
& \\  \hline

  \hline
{\footnotesize Jacobi structures  on  Lie algebra ${\rom{5}}$} &{\footnotesize   Equivalence classes}&
\\
 \hline
$ {\Lambda}_1 = 0
 $
&
$ \Lambda =0
 $
 &
\\
 ${E}_1= - e_{{1}}\partial _{x_1}-e_{{2}}\partial_{ x_2}
-e_{{3}}\partial _{x_3}
 $
&
 ${E}=- e_{{1}}\partial _{x_1}
  $
  &$e_{1} \in \mathbb{R}-\lbrace 0\rbrace$
 \\ \hline

$ {\Lambda}_2 = \dfrac {e_{{2}}\lambda_{{13}}}{e_{{3}}}\partial _{x_1}\wedge \partial_{ x_2}
+\lambda_{{13}}\partial _{x_1}\wedge \partial_{ x_3}+\lambda_
{{23}}\partial _{x_2}\wedge \partial_{ x_3}
 $
&
 $ \Lambda= \lambda_{{13}}\partial _{x_1}\wedge \partial_{ x_3}
 $
 &
 $\lambda_{13} \in \mathbb{R}-\lbrace 0\rbrace$
\\
 ${E}_2=-e_{{2}}\partial _{x_2}-e_{{3}}\partial _{x_3}
 $
&
${E}=-\partial _{x_3}
  $
  &
 \\ \hline

 $ {\Lambda}_3 =\lambda_{{12}}\partial _{x_1}\wedge \partial_{ x_2}+
\lambda_{{23}}\partial _{x_2}\wedge \partial_{ x_3}
 $
&
$
\Lambda=\lambda_{{12}}\partial _{x_1}\wedge \partial_{ x_2}
 $
 &
 $\lambda_{12} \in \mathbb{R}-\lbrace 0\rbrace$
\\
 ${E}_3= - e_{{2}}\partial _{x_2}
 $
&
 $E=-\partial_{ x_2}
  $
  &\\
     \hline

     \hline
{\footnotesize Jacobi structures  on  Lie algebra ${\rom{6}}_0$}
 & Equivalence classes&
\\
 \hline
 $ {\Lambda}_1 =
0
 $
&
 $ \Lambda =
0
$
&
\\
 ${E}_1= - e_{{1}}\partial _{x_1}-e_{{2}}\partial_{ x_2}
-e_{{3}}\partial _{x_3}
$
&
 $E=
 -e_{{3}}\partial_{ x_3}
 $
 &
 \\ \hline

${\Lambda}_2=
\lambda_{{12}}\partial _{x_1}\wedge \partial_{ x_2}+\lambda_{{13}}\partial _{x_1}\wedge \partial_{ x_3}+
\lambda_{{23}}\partial _{x_2}\wedge \partial_{ x_3}
$
&
$ \Lambda=\partial _{x_2}\wedge \partial_{ x_3}
$
&
 \\
 ${E}_2=\lambda_{{23}}\partial_{ x_1}+\lambda_
{{13}}\partial_{ x_2}
$
&
$
E= \partial_{ x_1}$
&
 \\  \hline

${\Lambda}_3=
\lambda_{{12}}\partial _{x_1}\wedge \partial_{ x_2}+\lambda_{{23}}\partial _{x_1}\wedge \partial_{ x_3}+
\lambda_{{23}}\partial _{x_2}\wedge \partial_{ x_3}
$
&
$ \Lambda=
\lambda_{{23}}\partial _{x_1}\wedge \partial_{ x_3}+
\lambda_{{23}}\partial _{x_2}\wedge \partial_{ x_3}
$&
 $\lambda_{23} \in \mathbb{R}-\lbrace 0\rbrace$
 \\
 ${E}_3= -e_{2}\partial_{ x_1}-e_{2}\partial_{ x_2}
$
&$E= -\partial_{ x_1}-\partial_{ x_2}
$
& \\  \hline

${\Lambda}_4=
 \lambda_{{12}}\partial _{x_1}\wedge \partial_{ x_2}
$
&
${\Lambda}=
 \lambda_{{12}}\partial _{x_1}\wedge \partial_{ x_2}
$
&
 $\lambda_{12} \in \mathbb{R}-\lbrace 0\rbrace$
 \\
 ${E}_4=   -e_{1}\partial_{ x_1}-e_{2}\partial_{ x_2}
$
&${E}=  -\partial_{ x_1}
$
&
 \\  \hline

     \end{tabular}

   \newpage
{\bf Table 4}:
(Continued.)\\
    \begin{tabular}{| l|l | l |  }
      \hline

      \hline
{\footnotesize Jacobi structures  on  Lie   algebra $ \rom{6}_a$} &{\footnotesize Equivalence classes  }&
\\
 \hline
 $ {\Lambda}_1 =
0
 $
&
 ${ \Lambda} =
0
$&
\\
 ${E}_1=- e_{{1}}\partial _{x_1}-e_{{2}}\partial_{ x_2}
-e_{{3}}\partial _{x_3}
$
&
 ${E}=
 - e_{{1}}\partial _{x_1}$
 &
  $e_{1} \in \mathbb{R}-\lbrace 0\rbrace$
 \\ \hline

${\Lambda}_2=
 \lambda_{{12}}\partial _{x_1}\wedge \partial_{ x_2}+\lambda_{{12}}\partial _{x_1}\wedge \partial_{ x_3}+
\lambda_{{23}}\partial _{x_2}\wedge \partial_{ x_3}$
&
${\Lambda}=
 \lambda_{{12}}\partial _{x_1}\wedge \partial_{ x_2}+\lambda_{{12}}\partial _{x_1}\wedge \partial_{ x_3}
$
&

 \\
 ${E}_2=
( a\lambda_{{12}}-
\lambda_{{12}})\partial _{x_2}+(a\lambda_{{12}}-\lambda_{{12}})\partial _{x_3}
$
&${E}=( a\lambda_{{12}}-
\lambda_{{12}})\partial _{x_2}+(a\lambda_{{12}}-\lambda_{{12}})\partial _{x_3}
$
& $\lambda_{12} \in \mathbb{R}-\lbrace 0\rbrace$
\\  \hline

${\Lambda}_3=
 \lambda_{{12}}\partial _{x_1}\wedge \partial_{ x_2}-\lambda_{{12}}\partial _{x_1}\wedge \partial_{ x_3}+
\lambda_{{23}}\partial _{x_2}\wedge \partial_{ x_3}
$
&
$\Lambda=
 \lambda_{{12}}\partial _{x_1}\wedge \partial_{ x_2}-\lambda_{{12}}\partial _{x_1}\wedge \partial_{ x_3}$
&
 $\lambda_{12} \in \mathbb{R}-\lbrace 0\rbrace$
 \\
 ${E}_3=-e_{{2}}\partial_{ x_2}
+e_{{2}}\partial _{x_3}
$
&$E= -\partial_{ x_2}
+\partial _{x_3}
$& \\  \hline

${\Lambda}_4=
 \lambda_{{12}}\partial _{x_1}\wedge \partial_{ x_2}+\lambda_{{12}}\partial _{x_1}\wedge \partial_{ x_3}+
\lambda_{{23}}\partial _{x_2}\wedge \partial_{ x_3}
$
&
$\Lambda=
 \lambda_{{12}}\partial _{x_1}\wedge \partial_{ x_2}+\lambda_{{12}}\partial _{x_1}\wedge \partial_{ x_3}$
&
 $\lambda_{12} \in \mathbb{R}-\lbrace 0\rbrace$
 \\
 ${E}_4=-e_{{2}}\partial_{ x_2}
-e_{{2}}\partial _{x_3}
$
&$E=-\partial_{ x_2}
-\partial _{x_3}
$&\\  \hline

${\Lambda}_5=
  \lambda_{{23}}\partial _{x_2}\wedge \partial_{ x_3}
$
&
$\Lambda=
 \lambda_{{23}}\partial _{x_2}\wedge \partial_{ x_3}
$
&
 $\lambda_{23} \in \mathbb{R}-\lbrace 0\rbrace$
 \\
 ${E}_5=-e_{{3}}\partial _{x_3}
$
&${E}=
-\partial _{x_3}
$
& \\  \hline

      \hline
{\footnotesize Jacobi structures  on  Lie algebra $ {\rom{7}_0}$ } &{\footnotesize Equivalence classes  }&
\\
 \hline
 $ {\Lambda}_1 =
0
 $
&
 $ \Lambda =
0
$&
\\
 ${E}_1=- e_{{1}}\partial _{x_1}-e_{{2}}\partial_{ x_2}
-e_{{3}}\partial _{x_3}
$
&
${E}=
-e_{{3}}\partial _{x_3}
 $& $e_{3} \in \mathbb{R}-\lbrace 0\rbrace$
 \\ \hline

${\Lambda}_2=
\lambda_{{12}}\partial _{x_1}\wedge \partial_{ x_2}+\lambda_{{13}}\partial _{x_1}\wedge \partial_{ x_3}+
\lambda_{{23}}\partial _{x_2}\wedge \partial_{ x_3}
$
&
$\Lambda=\partial _{x_2}\wedge \partial_{ x_3}
$&
$\lambda_{{13}}^2+\lambda_{{23}}^2\neq 0$
 \\
 ${E}_2= \lambda_{{23}}\partial _{x_1}-\lambda_
{{13}}\partial _{x_2}
$
&${E}=\partial _{x_1}
$& \\
 \hline

${\Lambda}_3=
 \lambda_{{12}}\partial _{x_1}\wedge \partial_{ x_2}
$
&
$
\Lambda=
 \lambda_{{12}}\partial _{x_1}\wedge \partial_{ x_2}
$
&
 $\lambda_{12} \in \mathbb{R}-\lbrace 0\rbrace$
 \\
 ${E}_3=  - e_{{1}}\partial _{x_1}-e_{{2}}\partial_{ x_2}
$
&${E}=-\partial_{ x_2}
$
&
\\  \hline

   \hline
{\footnotesize Jacobi structures  on  Lie algebra  $ \rom{7}_a$} &{\footnotesize Equivalence classes  }&
\\
 \hline
 $ {\Lambda}_1 = 0
 $
&
$\Lambda =0
 $&
\\
${E}_1= - e_{{1}}\partial _{x_1}-e_{{2}}\partial_{ x_2}
-e_{{3}}\partial _{x_3}
 $
&
 $E=- e_{{1}}\partial _{x_1}
  $& $e_{1} \in \mathbb{R}-\lbrace 0\rbrace$
 \\ \hline

${\Lambda}_2=
\lambda_{{23}}\partial _{x_2}\wedge \partial_{ x_3}
$
&
$\Lambda=
\lambda_{{23}}\partial _{x_2}\wedge \partial_{ x_3}
$
&
 $\lambda_{23} \in \mathbb{R}-\lbrace 0\rbrace$
 \\
 ${E}_2=-e_{{2}}\partial_{ x_2}
-e_{{3}}\partial _{x_3}
$
&${E}=
-\partial _{x_3}
$& \\  \hline

     \end{tabular}

   \section*{Appendix 3:  Automorphism Groups of Real  Low-Dimensional Lie  Algebras  }
{\bf Table 5}:
Automorphism groups of real two- and three-dimensional   Lie algebras ( see also \textcolor{red}{[\ref{Hemmati}]}).\\
\begin{tabular}{|l| l|  l|}
\hline

\hline
 { Lie Algebra } & { Automorphism groups} &{ Comments} \\
\hline
 {\footnotesize $ A_{1}$}& {\footnotesize $GL(2,\mathbb{R})$} & \\

 \hline
{\footnotesize $ A_{2}$}
 & {\footnotesize $\left(
\begin{array}{cc}
a_{11} & 0 \\
a_{21} & 1 \\
\end{array} \right)$} & {\footnotesize  $a_{11}\in\mathbb{R}-\{0\}$} \\

\hline
{\footnotesize $ I$}
& {\footnotesize $GL(3,R)$} &
\\

\hline
{\footnotesize $II$}
 & {\footnotesize $\left(
\begin{array}{ccc}
a_{22}a_{33}-a_{23}a_{32} & 0 & 0 \\
a_{21} &a_{22} &a_{23} \\
a_{31} &a_{32} &a_{33}
\end{array} \right)$} &
{\footnotesize   $ a_{21},a_{22},a_{23},
a_{31},a_{32},a_{33}\in\mathbb{R},  a_{22}a_{33} \neq a_{23}a_{32}$}
  \\
\hline
 {\footnotesize $III, VI_{a}$}
 & {\footnotesize $\left(
\begin{array}{ccc}
1 & a_{12} & a_{13} \\
0 & a_{22} & a_{23} \\
0 &  a_{23} &a_{22}
\end{array} \right)$} &
{\footnotesize  $ a_{12}, a_{13},
 a_{22}, a_{23}\in\mathbb{R},  a_{22}\neq \pm a_{23}$}
 \\
\hline
 {\footnotesize $IV$}
& {\footnotesize $\left(
\begin{array}{ccc}
1 & a_{12} & a_{13} \\
0 & a_{22} & a_{23} \\
0 &  0 &a_{22}
\end{array} \right)$} &
{\footnotesize   $ a_{12}, a_{13},
 a_{23}\in\mathbb{R},   a_{22}\in\mathbb{R}-\{0\}$}
\\
\hline
 {\footnotesize $V$}
 & {\footnotesize $\left(
\begin{array}{ccc}
1 & a_{12} & a_{13} \\
0 & a_{22} & a_{23} \\
0 &  a_{32} &a_{33}
\end{array} \right)$} &
{\footnotesize  $ a_{12},a_{13},a_{22},a_{23},
a_{32},a_{33}\in\mathbb{R},  a_{22}a_{33} \neq a_{23}a_{32}$}
\\
\hline
{\footnotesize $VI_{0}$}
 & {\footnotesize $\left(
\begin{array}{ccc}
a_{11} & a_{12} & 0 \\
a_{12} & a_{11} & 0 \\
a_{31} & a_{32} & 1
\end{array} \right)$} $,$ {\footnotesize $\left(
\begin{array}{ccc}
a_{11} & a_{12} & 0 \\
-a_{12} & -a_{11} & 0 \\
a_{31} & a_{32} & -1
\end{array} \right)$}&
{\footnotesize $ a_{11},a_{12},a_{31},a_{32}\in\mathbb{R},  a_{11} \neq \pm a_{12}$}
  \\
\hline

{\footnotesize $VII_{0}$}
& {\footnotesize $\left(
\begin{array}{ccc}
a_{11} & a_{12} & 0 \\
-a_{12} & a_{11} & 0 \\
a_{31} & a_{32} & 1
\end{array} \right)$} $,$ {\footnotesize $\left(
\begin{array}{ccc}
a_{11} & a_{12} & 0 \\
a_{12} & -a_{11} & 0 \\
a_{31} & a_{32} & -1
\end{array} \right)$} &
{\footnotesize $   a_{11},a_{12},a_{31},a_{32}\in\mathbb{R},   a_{11}^{2}+a_{12}^{2}\neq 0$}
 \\
\hline

 {\footnotesize $VII_{a}$}
 & {\footnotesize $\left(
\begin{array}{ccc}
1 & a_{12} &  a_{13} \\
0 &  a_{22} & - a_{23} \\
0 & a_{23} & a_{22}
\end{array} \right)$} &
{\footnotesize $a_{12},a_{13},a_{22},a_{23}\in\mathbb{R},   a_{22}^{2}+a_{23}^{2}\neq 0$}
 \\
\hline
{\footnotesize $VIII$}
&  {\footnotesize $SL(2,\mathbb{R})$}  &  \\
\hline
{\footnotesize $IX$}
&  {\footnotesize $SO(3)$} &  \\
\hline
\end{tabular}

        \newpage

\end{document}